\documentclass[preprint]{iucr}
\RequirePackage{graphicx}
\usepackage{amsmath}           
\usepackage{allrunes}
\usepackage{rotating}
\usepackage{mathtools}
\usepackage{upgreek}
\usepackage{dsfont}
\usepackage{rotating}

\usepackage{xcolor}

\newcommand{\raidob}{\textarc{\textbf{r}}}

\newcommand{\ehwaz}{\textarc{e}}

\DeclareMathOperator{\acos}{acos}

\DeclareMathOperator{\atan}{atan}

\usepackage{url}

\usepackage[top=2cm,left=2.5cm,right=2.5cm,bottom=2.5cm]{geometry}
     \journalcode{M}              

\begin{document}                  



\title{General method to calculate the elastic deformation and X-ray diffraction properties of bent crystal wafers}


\cauthor{Ari-Pekka}{Honkanen\footnote[1]{\includegraphics[scale=0.75]{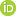} \url{https://orcid.org/0000-0002-6822-3062}}}{ari-pekka.honkanen@helsinki.fi}{}
\author{Simo}{Huotari\footnote[2]{\includegraphics[scale=0.75]{orcid.png} \url{https://orcid.org/0000-0003-4506-8722}}}

\aff{University of Helsinki, Department of Physics, PO Box 64, FI-00014 Helsinki, \country{Finland}}









\maketitle                        


\begin{abstract}
Curved single crystals are widely employed in spectrometer designs in the hard X-ray regime. Due to their large solid angle coverage and focusing properties, toroidally bent crystals are extremely useful in applications where the output of photons is low. Spherically bent crystals, a subgroup of toroidally bent crystals, particularly have found their way in many instruments at synchtrotrons and free electron laser lightsource end-stations but also in the re-emerging field of high-resolution laboratory-scale X-ray spectroscopy. A solid theoretical understanding of the diffraction properties of such crystals is essential when aiming for optimal spectrometer performance. In this work, we present a general method to calculate the internal stress and strain fields of toroidally bent crystals and how to apply it to predict their diffraction properties. Solutions are derived and discussed for circular and rectangular spherically bent wafers due to their prevalence in contemporary instrumentation.
\end{abstract}
\vfill
\pagebreak


\section{Introduction}

Crystal analysers are the heart of most contemporary mid-to-high energy resolution X-ray spectrometers in the hard X-ray regime. The same basic principle, the diffraction of X-rays from the periodical crystal structure, has conceived a plethora of spectrometric designs, of which many employ curved crystal analysers to increase the flux of collected photons and to ensure their proper focusing on a detector \cite{DuMond_1930, Johann_1931, Johansson_1932,Cauchois_1932,von_Hamos_1932}. 
Especially with spherically bent crystal analysers (SBCA) one can efficiently cover and analyse photons collected over a large solid angle. SBCAs also exhibit (approximate) point-to-point focusing allowing integration of imaging and tomography capabilities in to spectroscopic instruments \cite{Huotari_2011}. It is no wonder that many inelastic X-ray scattering (IXS) and X-ray emission spectroscopy (XES) end-stations at synchrotron and free electron laser lightsources worldwide, such as SOLEIL \cite{Ablett_2019}, ESRF \cite{Kvashnina_2016,Huotari_2017, Moretti_Sala_2018}, APS \cite{Fister_2006}, Spring-8 \cite{Cai_2004,Ishii_2013}, SSRF \cite{Duan_2016}, SLS \cite{Kleymenov_2011}, SSRL \cite{Sokaras_2012}, and DESY \cite{Welter_2005}, utilize SBCAs in their instrument designs. In addition to studying the structure and internal dynamics of matter \emph{via} externally produced radition, SBCAs are also used to analyse X-rays in plasma research \cite{Faenov_1994,Aglitskiy_1998,Sinars_2003,Knapp_2011}.

Due to high demand and limitations of synchrotron/free electron access, a renewed interest toward laboratory-scale X-ray instrumentation based on conventional X-ray tubes has grown in recent years  \cite{Seidler_2014,Anklamm_2014,Nemeth_2016,Holden_2017,Honkanen_2019,Jahrman_2019}. 
Especially relevant to this work are the instrument designs based on SBCAs which, in conjunction with recent advances in the crystal technology \cite{Verbeni_2005,Rovezzi_2017}, have largely overcome the problem of low photon output plagueing the previous generation of laboratory instruments (that were often based on cylindrically bent crystals) \cite{Seidler_2014}. 
Indeed, the portfolio of scientific cases, in which the lab instruments using SBCAs have proven to be a viable alternative to large-scale facilities, is expanding rapidly and spans already a vast cavalcade of interests in natural sciences such as fundamental materials research \cite{Mortensen_2017}, electrochemistry \cite{Wang_2017, Kuai_2018, Sun_2019, Lutz_2020}, nanoparticle characterisation \cite{Davodi_2019}, \emph{in operando} battery studies \cite{Jahrman_2018,Jahrman_2019c}, actinide research \cite{Bes_2018, Jahrman_2019,Mottram_2020}, \emph{in situ} catalysis studies \cite{Moya_Cancino_2019,Moya_Cancino_2019b}, geochemistry \cite{Mottram_2020b}, and microbiology and enviromental research \cite{Lusa_2019}.

However, as a significant disadvantage SBCAs suffer from focal astigmatism when taken out of the backscattering condition which can cause aberrations in imaging and issues with detectors with small active areas. The problem can be averted with toroidally bent crystal analysers (TBCA) which have different sagittal and meridional bending radii. Notwithstanding, TBCAs are encountered rarely as they are more difficult to manufacture than SBCAs and need to be tuned for a specific Bragg angle which incurs increased expenses, especially if the spectrometer setup is meant to be used for a wide range of photon energies. However, at least some of these problems can be avoided by using vacuum-forming optics \cite{Jahrman_2019b} to apply the toroidal bending to a flat wafer temporarily and, perhaps with further development, dynamically in the course of an experiment.

In general, the bending process degrades the energy resolution of a TBCA/SBCA by introducing internal stress to the crystal wafer. The effect can be mitigated \emph{e.g.} by dicing or cutting the crystal surface \cite{Verbeni_2005,Verbeni_2009,Shvyd_ko_2013}. However, without a guiding theoretical understanding, such mechanical alterations might lead to unexpected adverse effects, such as loss of integrated reflectivity, optical aberrations, and increased manufacturing costs. From the standpoint of instrument optimization it is thus of utmost importance to understand how the diffractive properties and the mechanical deformation of toroidally/spherically bent crystal wafer are intertwined together.  

The equations describing the propagation of radiation in deformed periodic medium were laid out independently by S. Takagi and D. Taupin in 1960s \cite{Takagi_1962,Taupin_1964,Takagi_1969} which together with lamellar models \cite{White_1950,Erola_1990,Rio_2004} are routinely used to calculate the diffraction properties of bent crystals \cite{Gronkowski_1991,Sanchez_del_Rio_2011}. However, an adequate theory to calculate the internal strains inside a spherically bent crystal wafer and thus its diffraction properties were lacking until mid-2010s \cite{honkanen_14,Honkanen_2014b,Honkanen_2016}. Inclusion of in-plane strains to a thin wafer \emph{via} geometrical considerations and anisotropic linear elasticity leads to a model that can accuraterly predict the experimentally measured reflectivity curves of SBCAs with circularly shaped wafers cut along arbitrary crystal directions. Nevertheless, the original derivation relies on many geometrical features and symmetries which can not be easily generalized to toroidal bending or other types of crystal shapes, such as rectangular ones used \emph{e.g.} in recently introduced strip-bent analysers \cite{Rovezzi_2017}.

In this work, we present a general framework to calculate internal stress and strain fields and diffraction curves of an arbitrarily shaped, toroidally bent crystal wafer. The procedure is utilized to derive stress and strain expressions for isotropic and anisotropic circular and rectangular spherically bent crystals due to their prevalence in the contemporary instrumentation scene. The models and their properties are discussed in detail and the accuracy of the predicted diffraction curves is validated by comparison to experimental data. The Python implementation of the models is briefly introduced.

\section{Theory\label{sec:theory}}

The propagation of the electromagnetic radiation in deformed medium is mathematically described by a group of partial differential equations known as the Takagi-Taupin equations \cite{Takagi_1962,Taupin_1964,Takagi_1969}. To accurately compute a diffraction curve of a bent crystal, the strain tensor needs to be known over the diffraction domain. In what follows, a general procedure to obtain the deformation field of a toroidally bent, thin anisotropic crystal wafer is presented.

\subsection{Solving the deformation field of arbitrarily shaped toroidally bent crystal wafer}

Consider a thin anisotropic crystal wafer of thickness $d$. We choose a Cartesian coordinate system $(x,y,z)$ so that the origin of the system coincides with the midplane of the wafer with the $z$-direction parallel to the normal of the crystal surface. The displacement vector field $\boldsymbol{\epsilon}$ due to two orthogonal torques acting on the wafer about the $x$- and $y$-axes is \cite{Chukhovskii_1994}
\begin{align}
\epsilon_x &= (S_{11} \mu_x + S_{12} \mu_y) x z + (S_{51} \mu_x + S_{52} \mu_y)\frac{z^2}{2} + (S_{61} \mu_x +S_{62} \mu_y) \frac{y z}{2} \label{eq:pure_bending_x} \\
\epsilon_y &= (S_{21} \mu_x + S_{22} \mu_y) y z + (S_{41} \mu_x + S_{42} \mu_y)\frac{z^2}{2} + (S_{61} \mu_x +S_{62} \mu_y) \frac{x z}{2} 
\label{eq:pure_bending_y}\\
\epsilon_z &= -(S_{11} \mu_x + S_{12} \mu_y)\frac{x^2}{2} -(S_{21} \mu_x + S_{22} \mu_y)\frac{y^2}{2} \nonumber \\ &\quad -(S_{61} \mu_x +S_{62} \mu_y) \frac{x y}{2} +(S_{31} \mu_x + S_{32} \mu_y)\frac{z^2}{2} \label{eq:pure_bending_z}
\end{align} 
where $S_{ij}$ are components of the compliance matrix as used in the Voigt notation\footnote{In the Voigt notation, a pair of indices $ij$ is replaced with a single index $m$ as follows: $11 \rightarrow 1$; $22 \rightarrow 2$; $33 \rightarrow 3$; $23,32 \rightarrow 4$; $13, 31 \rightarrow 5$ and $12, 21 \rightarrow 6$. The compliance matrix $S$ in the Voigt notation is given in terms of the compliance tensor $s$ so that $S_{mn} = (2 - \delta_{ij})(2 - \delta_{kl})s_{ijkl}$, where $ij$ and $kl$ are any pairs of indices corresponding to $m$ and $n$, respectively, and $\delta$ is the Kronecker delta.}
The torques $\mu_x$ and $\mu_y$ are in units of torque per unit length per the area moment of inertia. The subscript of the scaled torques refers to direction along which the torque primarily bends the crystal, not their axes ($\mu_x$ acts about the $y$-axis and $\mu_y$ about the $x$-axis). From the form of Eq.~\eqref{eq:pure_bending_z} we see that the torques cause the wafer to deform into the shape of a paraboloid approximating well the toroidal shape when the dimensions of the wafer are small compared to the radii of curvature.

The displacement vector field \eqref{eq:pure_bending_x}--\eqref{eq:pure_bending_z} applies for the case where the deformation is sufficiently small to not cause significant streching in the in-plane directions and is thus called a \emph{pure bending solution}. By imposing the requirement that the midplane ($z=0$) of the wafer needs to follow the shape of the paraboloid surface \emph{i.e.}
\begin{equation}
\epsilon_z(x,y,0) = \left(\frac{\cos^2 \phi}{R_1} + \frac{\sin^2 \phi}{R_2}\right)\frac{x^2}{2} -  \sin 2\phi \left(\frac{1}{R_1} -  \frac{1}{R_2}\right) \frac{xy}{2} +
\left(\frac{\sin^2 \phi}{R_1} + \frac{\cos^2 \phi}{R_2}\right)\frac{y^2}{2} \label{eq:general_toroidal_deflection}
\end{equation}
where $R_1$ and $R_2$ are the radii of curvature and $\phi$ is the in-plane inclination of the main axis of curvature with the coordinate system (clockwise-positive), one could in principle solve the $\mu_x$ and $\mu_y$ required to produce the sougth-after deflection profile. The obtained deformation field can be used to solve the diffraction curve of a wafer with small enough surface area to not be influenced by the transverse streching and thus the shape of the wafer.

However, since there are two torques and three parameters that define the shape and orientation of the deflection in $z$, the only two of $R_1$, $R_2$, and $\phi$ can be chosen freely and the third one is determined by $S_{ij}$. For example, in the case of spherical bending $R_1 = R_2$ which means that the $xy$-term in Eq.~\eqref{eq:general_toroidal_deflection} should vanish. However, in general for non-zero $S_{61}$ and $S_{62}$ there are no $\mu_x$ and $\mu_y$ acting about the arbitrarily fixed cardinal axes that would equate the displacement vector in $z$ with the spherical surface. 

The torques acting on the wafer \emph{in natura} are imposed by the contact to the substrate onto which the wafer is forced and can choose their axes of action freely to conform the shape of the wafer to that of the substrate. The solution \eqref{eq:pure_bending_x}--\eqref{eq:pure_bending_z} assumes that $\mu_x$ and $\mu_y$ act about fixed axes but mathematically the same effect can be achieved by introducing an additional rotational degree of freedom $\alpha$ to the crystal directions and $\phi$ in \emph{xy}-plane. Combining Eqs.~\eqref{eq:pure_bending_z} and \eqref{eq:general_toroidal_deflection} with well-known trigonometric identities, we thus need to find the torques $\mu_x$ and $\mu_y$ and the in-plane rotation angle $\alpha$ so that the following equations are fulfilled simultaneously: 
\begin{align}
S_{11}' \mu_x + S_{12}' \mu_y &= - \frac{1}{2} \left(\frac{1}{R_1} + \frac{1}{R_2} \right) - \frac{1}{2} \left(\frac{1}{R_1} - \frac{1}{R_2} \right)\cos 2 \phi' \label{eq:toroidal_condition1} \\
S_{21}' \mu_x + S_{22}' \mu_y &= - \frac{1}{2} \left(\frac{1}{R_1} + \frac{1}{R_2} \right) + \frac{1}{2} \left(\frac{1}{R_1} - \frac{1}{R_2} \right)\cos 2 \phi' \label{eq:toroidal_condition2} \\
S_{61}' \mu_x + S_{62}' \mu_y &=  \left(\frac{1}{R_1} - \frac{1}{R_2} \right)\sin 2\phi' \label{eq:toroidal_condition3}
\end{align}
where $S_{ij}'$ are the components of the rotated compliance matrix and $\phi' = \phi + \alpha$. Without a loss of generality, we may assume that non-rotated $S_{ij}$ are originally presented in a coordinate system that is parallel with the main axes of the toroidal bending thus allowing us to set $\phi = 0$. Using the first two of the equations we find that
\begin{align}
\mu_x = \frac{(S_{12}' - S_{22}')(R_1 + R_2) + (S_{12}' + S_{22}')(R_1 - R_2)\cos 2 \alpha}{2(S_{11}'S_{22}'  - S_{12}'S_{12}')R_1 R_2} \label{eq:torque_x}\\
\mu_y = \frac{(S_{12}' - S_{11}')(R_1 + R_2) - (S_{12}' + S_{11}')(R_1 - R_2)\cos 2 \alpha}{2(S_{11}'S_{22}'  - S_{12}'S_{12}')R_1 R_2} \label{eq:torque_y}
\end{align}
where $S_{12}'=S_{21}'$ based on the symmetry of $S$ was used. Now, substituting the obtained torques to Equation~\eqref{eq:toroidal_condition3} leads to the condition
\begin{align}
\Big[ 2 (S_{12}'S_{12}'-S_{11}'S_{22}' ) \sin 2 \alpha &+  
\left[S_{61}' (S_{22}' + S_{12}') - S_{62}' (S_{11}' +S_{12}' ) \right]\cos 2 \alpha   \Big] (R_1  - R_2 ) \nonumber \\
&= \left[S_{61}' (S_{22}' - S_{12}') + S_{62}' (S_{11}' - S_{12}') \right](R_1  + R_2 ) \label{eq:alpha_condition}
\end{align}
The in-plane rotation angle $\alpha$ fulfilling the condition~\eqref{eq:alpha_condition} can be solved by performing a rotation to the compliance tensor $s$ according to
\begin{equation}
s'_{ijkl} = \sum_{p,q,r,s} Q_{ip}Q_{jq}Q_{kr}Q_{ls} s_{pqrs} \label{eq:tensor_rotation}
\end{equation}
where $Q$ is the rotation matrix corresponding to the counterclockwise rotation by $\alpha$ about $z$-axis that is given by 
\begin{equation}
Q = \left[\begin{matrix}
\cos \alpha & -\sin \alpha & 0 \\
\sin \alpha & \cos \alpha & 0 \\
0 & 0 & 1
\end{matrix}\right].
\end{equation} 
Constructing the relevant components of the rotated compliance matrix $S'$ from $s_{ijkl}'$ allows us to write the Eq.~\eqref{eq:alpha_condition} in terms of $S$:
\begin{equation}
( A_\alpha \sin 2 \alpha + B_\alpha \cos 2 \alpha ) (R_1 - R_2)
= ( C_\alpha \sin 2 \alpha + D_\alpha \cos 2 \alpha ) (R_1 + R_2)
\end{equation}
where
\begin{align}
A_\alpha &\equiv S_{66} (S_{11} + S_{22} + 2 S_{12}) - (S_{61} + S_{62})^2 \\
B_\alpha &\equiv  2\left[S_{62} (S_{12} + S_{11}) - S_{61} (S_{12} + S_{22}) \right]   \\
C_\alpha &\equiv S_{66} (S_{22} - S_{11}) + S_{61}^2 - S_{62}^2  \\
D_\alpha &\equiv 2\left[S_{62} (S_{12} - S_{11}) + S_{61} (S_{12} - S_{22}) \right].
\end{align}
Solving for $\alpha$, we find
\begin{equation}
\alpha = \frac{1}{2} \atan \left[ \frac{D_\alpha(R_1 + R_2) - B_\alpha (R_1 - R_2)}{ A_\alpha (R_1 - R_2) - C_\alpha (R_1 + R_2) } \right] + \frac{\pi n}{2}, \label{eq:alpha_angle_toroidal}
\end{equation}
where $n \in \mathds{Z}$. The derivation of the obtained expression is based on the assumption that at least either of $S'_{61}$ or $S'_{62}$ is non-zero. By examining the rotated components in detail, we find that this assumption fails if the following conditions are simultaneously true: $S_{61} = S_{62} = 0$, $S_{11} = S_{22}$, and $S_{11} + S_{22} - 2 S_{12} - S_{66} = 0$. Elastically isotropic material, for example, fulfils these conditions. In such a case, Eq~\eqref{eq:toroidal_condition3} reduces to $\sin 2\alpha = 0$ which leads to $\alpha = \pi n /2$. Since any valid $\alpha$ suits the purpose, we may choose $n=0$ for simplicity in both cases.

Since the crystal does not rotate physically, we need to compensate the tensor rotation by rotating the coordinate system with it. This means the rotation of the displacement vector $\boldsymbol{\epsilon}' = Q^T \boldsymbol{\epsilon}$ and replacement of the scalar coordinates by $x \rightarrow x\cos \alpha + y \sin \alpha$ and $y \rightarrow y \cos \alpha - x \sin \alpha$.\footnote{Note that we do not apply the rotation to the compliance tensor in $S_{ij}'\mu_k$ as these products behave as scalars.} Thus the components of the displacement vector field in the pure bending solution [Eqs.~\eqref{eq:pure_bending_x}--\eqref{eq:pure_bending_z}] for spherical bending become
\begin{align}
\epsilon_x' &= -\frac{xz}{R_1} +\left[(S_{51}'\mu_x + S_{52}' \mu_y ) \cos \alpha - (S_{41}'\mu_x +S_{42}' \mu_y ) \sin \alpha  \right] \frac{z^2}{2} \\
\epsilon_y' &= -\frac{yz}{R_2}
+\left[(S_{51}'\mu_x + S_{52}' \mu_y ) \sin \alpha + (S_{41}'\mu_x +S_{42}' \mu_y ) \cos \alpha  \right] \frac{z^2}{2} \\
\epsilon_z' &= \frac{x^2}{2 R_1} + \frac{y^2}{2R_2}
+ (S_{31}'\mu_x + S_{32}'\mu_y)\frac{z^2}{2}
\end{align}
where the $S_{ij}'$, $\alpha$, $\mu_x$ and $\mu_y$ are best calculated numerically using Eqs.~\eqref{eq:torque_x}, \eqref{eq:torque_y}, \eqref{eq:tensor_rotation} and \eqref{eq:alpha_angle_toroidal}. Assuming the diffraction to take place in the $xz$-plane, the partial derivatives needed for the diffraction calculations are thus found to be
\begin{align}
&\frac{\partial \epsilon_x'}{\partial x} = -\frac{z}{R_1} \qquad \quad
\frac{\partial \epsilon_z'}{\partial x} = \frac{x}{R_1} 
\qquad \quad
\frac{\partial \epsilon_z'}{\partial z} = (S_{31}' \mu_x + S_{32}' \mu_y)z
\nonumber \\
&\frac{\partial \epsilon_x'}{\partial z}
= -\frac{x}{R_1}
+\left[(S_{51}'\mu_x + S_{52}' \mu_y ) \cos \alpha - (S_{41}'\mu_x +S_{42}' \mu_y ) \sin \alpha  \right] z \label{eq:pure_bending_partial_derivatives_toroidal}
\end{align}
In the isotropic case\footnote{The non-zero components are $S_{11}' = S_{22}' = S_{33}' = 1/E$, $S_{12}' = S_{21}' = S_{13}' = S_{31}'= S_{23}' = S_{32}' = -\nu/E$, and $S_{44}' = S_{55}' = S_{66}' = 2(1+\nu)/E$}, the torques given by Eqs.~\eqref{eq:torque_x} and \eqref{eq:torque_y} reduce to
\begin{equation}
\mu_x = -\frac{E}{1-\nu^2}\left( \frac{1}{R_1} + \frac{\nu}{R_2} \right) \qquad
\mu_y = -\frac{E}{1-\nu^2}\left( \frac{\nu}{R_1} + \frac{1}{R_2} \right)
\end{equation}
and thus the partial derivatives of $\boldsymbol{\epsilon}'$ become
\begin{equation}
\frac{\partial \epsilon_x'}{\partial x} = -\frac{z}{R_1} \qquad
\frac{\partial \epsilon_x'}{\partial z} = -\frac{x}{R_1} \qquad 
\frac{\partial \epsilon_z'}{\partial x} = \frac{x}{R_1}  \qquad
\frac{\partial \epsilon_z'}{\partial z} = \frac{\nu}{1-\nu} \left(\frac{1}{R_1} + \frac{1}{R_2} \right)z
\end{equation}

The partial derivative \eqref{eq:pure_bending_partial_derivatives_toroidal} can be used as a deformation term in the Takagi-Taupin equations to estimate X-ray diffraction curves of toroidally bent crystals. However, the pure bending solution alone is inadequate as it fails to explain the resolution function of SBCAs with a large surface area \cite{Verbeni_2009, honkanen_14, Rovezzi_2017}. This is because, in addition to pure bending strain, the flat crystal wafer is also stretched and compressed in the transverse directions in order to fit on a spherical surface. These deformations affect the separation of the diffracting Bragg planes due to non-zero Poisson ratio and thus the resolution function of the SBCA. In the scope of linear elasticity, the total strain tensor is $\tilde{\epsilon}_{ij} = \epsilon_{ij}' + u_{ij}$, where in addition to the pure bending strain $\epsilon_{ij}'$ we include the \emph{streching component} $u_{ij}$. In what follows, a theoretical foundation for solving $u_{ij}$ is presented.

According to Hooke's law, the components of the strain tensor due to stretching $u_{ij}$ are connected to the stretching stress tensor $\sigma_{ij}$ \emph{via}   
\begin{equation}\label{eq:Hookes_law}
u_{ij} = \sum_{k,l}s_{ijkl}\sigma_{kl}
\end{equation} where $s_{ijkl}$ is the compliance tensor.
Using the Voigt notation to convert the fourth-order compliance tensor to a matrix, Equation~\eqref{eq:Hookes_law} gives the following relations
\begin{align}
u_{xx} &= S_{11} \sigma_{xx} + S_{12} \sigma_{yy} + S_{16} \sigma_{xy} \label{eq:aniso_uxx}\\
u_{yy} &= S_{21} \sigma_{xx} + S_{22} \sigma_{yy} + S_{26} \sigma_{xy} \label{eq:aniso_uyy} \\
u_{xy} &= \frac{1}{2}\left(S_{61}\sigma_{xx} + S_{62} \sigma_{yy} + S_{66} \sigma_{xy} \right). \label{eq:aniso_uxy}
\end{align}
In Eqs.\eqref{eq:aniso_uxx}--\eqref{eq:aniso_uxy} we have assumed $\sigma_{xz}=\sigma_{yz}=\sigma_{zz} = 0$, since the external forces required to bend a thin plate are small compared to the internal stresses and can thus be omitted at this stage. For an isotropic crystal, the relations simplify to 
\begin{equation}
u_{xx} = \frac{\sigma_{xx}-\nu \sigma_{yy}}{E} \qquad u_{yy} = \frac{\sigma_{yy}-\nu \sigma_{xx}}{E} \qquad u_{xy} = \frac{1+\nu}{E}\sigma_{xy},\label{eq:lateral_strains}
\end{equation}
where $E$ is Young's modulus and $\nu$ is Poisson's ratio.

The transverse components of $u_{ij}$ are given by Eq.~(14.1) in \cite{landau_lifshitz}[p. 51]
as follows
\begin{equation}
u_{ij} = \frac{1}{2}\left(\frac{\partial u_i}{\partial x_j} + \frac{\partial u_j}{\partial x_i} \right) + \frac{1}{2}\frac{\partial \zeta}{\partial x_i}\frac{\partial \zeta}{\partial x_j}, \label{eq:uij_large_deflection}
\end{equation}
where $u_i$ are the components of the displacement vector due to stretching and $\zeta$ is the vertical displacement of the wafer. The possible values of $i$ and $j$ are now restricted to the in-plane directions $x$ and $y$. The strain tensor must fulfill the equilibrium condition $\sum_k \partial \sigma_{ik} / \partial {x}_k = 0$ which is ascertained if we write the $\sigma_{ij}$ as a function of $\chi = \chi(x,y)$, also known as the Airy stress function, so that 
\begin{equation}
\sigma_{xx} = \frac{\partial^2 \chi}{\partial y^2}, \qquad
\sigma_{xy} = -\frac{\partial^2 \chi}{\partial x \partial y}, \qquad
\sigma_{yy} = \frac{\partial^2 \chi}{\partial x^2}. \label{eq:stresses}
\end{equation}

We are now set to find $u_{ij}$ which we will achieve by minimising the relevant thermodynamic potential, that is, the Helmholtz energy. The Helmholtz energy for the mechanical deformation of a thin wafer can be written as the sum of the pure bending and stretching energies. However, since the pure bending solution is already assumed to be known [Eqs.~\eqref{eq:pure_bending_x}--\eqref{eq:pure_bending_z}], we may focus only on the streching part given by
\begin{equation}
\mathcal{F} = \frac{d}{2} \int_\Omega  d\Omega \sum_{k,l} u_{kl} \sigma_{kl} = \frac{d}{2} \int_\Omega  d\Omega \ \Big( u_{xx} \sigma_{xx} + 2 u_{xy} \sigma_{xy}
+ u_{yy} \sigma_{yy} \Big) ,
\end{equation}
where the integration goes over the crystal surface $\Omega$. Substituting Eqs.\eqref{eq:aniso_uxx}--\eqref{eq:aniso_uxy}, we obtain
\begin{equation}
\mathcal{F} = \frac{d}{2} \int_\Omega  d\Omega \ \Big( S_{11} \sigma_{xx}^2 + S_{22} \sigma_{yy}^2 + S_{66} \sigma_{xy}^2
+ 2 S_{12} \sigma_{xx} \sigma_{yy} + 2 S_{16} \sigma_{xx} \sigma_{xy}
+ 2 S_{26} \sigma_{yy} \sigma_{xy} \Big), \label{eq:F_anisotropic}
\end{equation}
which in the isotropic case simplifies to
\begin{equation}
\mathcal{F} = \frac{d}{2 E} \int_\Omega  d\Omega \ \Big[ \sigma_{xx}^2 + 2(1+\nu) \sigma_{xy}^2 + \sigma_{yy}^2 - 2 \nu \sigma_{xx}\sigma_{yy} \Big].
\label{eq:F_isotropic}
\end{equation}

The deformation field is can be now found by minimizing $\mathcal{F}$ in terms of $\chi$, \emph{i.e.,} we need to find $\chi$ so that the functional derivative $\delta \mathcal{F}/\delta \chi = 0$. While we could try to solve the problem using the Euler-Lagrange equations, we may utilize the fact that the dimensions of the crystals are small compared to the bending radii $R_{1,2}$. Therefore we may write the ansatz in powers of $x/R_{1,2}$ and $y/R_{1,2}$ and truncate the series after a few lowest-order terms. The $\mathcal{F}$ is then minimized in terms of the expansion coefficients $C_k$. Since $\mathcal{F}$ is quadratic in terms of $\chi$ and thus in terms of $C_k$, the problem of solving the Euler-Lagrange equations is thus reduced to a finite linear system $\partial \mathcal{F}/\partial C_k = 0$. Taking the partial derivatives of Eq.~\eqref{eq:F_anisotropic}, we find
\begin{align}
\partial_k \mathcal{F} =  d \int_\Omega  d\Omega \ \Big[ &\left(S_{11} \partial_k \sigma_{xx} + S_{12} \partial_k \sigma_{yy} + S_{16} \partial_k \sigma_{xy} \right) \sigma_{xx} \nonumber \\ + &\left(S_{12} \partial_k \sigma_{xx} + S_{22} \partial_k \sigma_{yy} + S_{26} \partial_k \sigma_{xy} \right) \sigma_{yy} \nonumber \\ + &\left(S_{16} \partial_k \sigma_{xx} + S_{26} \partial_k \sigma_{yy} + S_{66} \partial_k \sigma_{xy} \right) \sigma_{xy} \Big] \label{eq:F_derivative_anisotropic}
\end{align}
where a shorthand $\partial_k \equiv \partial/\partial C_k$ has been used. For the isotropic crystal the equations simplify to
\begin{equation}
\partial_k \mathcal{F} =  \frac{d}{E} \int_\Omega  d\Omega \  \Big[
\left(\partial_k \sigma_{xx} - \nu \partial_k \sigma_{yy} \right) \sigma_{xx}
+ \left(\partial_k \sigma_{yy} - \nu \partial_k \sigma_{xx} \right) \sigma_{yy}
+ 2 (1 + \nu)( \partial_k \sigma_{xy}) \sigma_{xy} \Big]. \label{eq:F_derivative}
\end{equation}

In addition, we need to impose two constraints to the energy minimization to include the toroidal bending and the requirement that the integrated contact force at the wafer--substrate interface acting on the wafer vanishes. First, for the toroidal bending we need to find the relationship between $\chi$ and the vertical displacement $\zeta$. As presented in Appendix~\ref{app:lateral_strain}, by combining Eqs.~\eqref{eq:aniso_uxx}--\eqref{eq:aniso_uxy}, \eqref{eq:uij_large_deflection}, and \eqref{eq:stresses}, we obtain the following partial differential equation
\begin{equation}\label{eq:chi_zeta_relation}
\mathcal{D}^4\chi = \left(\frac{\partial^2 \zeta}{\partial x \partial y}\right)^2 -\frac{\partial^2 \zeta}{\partial x^2}\frac{\partial^2 \zeta}{\partial y^2},
\end{equation} 
where
\begin{equation}
\mathcal{D}^4 \equiv S_{11} \frac{\partial^4}{\partial y^4} + (2 S_{12} + S_{66}) \frac{\partial^4}{\partial x^2 \partial y^2} +  S_{22} \frac{\partial^4}{\partial x^4}
-2 S_{16}\frac{\partial^4}{\partial x \partial y^3} - 2 S_{26} \frac{\partial^4}{\partial x^3 \partial y}.
\end{equation}
Substituting the toroidal displacement $\zeta(x,y) = x^2/2R_1 +y^2/2R_2$ into Eq.~\eqref{eq:chi_zeta_relation}, we thus obtain
\begin{equation}\label{eq:chi_zeta_relation_aniso_toroidal}
\mathcal{D}^4\chi = -\frac{1}{R_1 R_2},
\end{equation}
which in the isotropic case simplifies to
\begin{equation}\label{eq:chi_zeta_relation_iso_toroidal}
\nabla^4\chi = -\frac{E}{R_1 R_2}.
\end{equation}

Second, as given in Appendix~\ref{app:contact_force}, the contact force $P$ per unit area acting on the wafer at the wafer--substrate interface is
\begin{equation}
P = -d \left(\frac{\sigma_{xx}}{R_1} + \frac{\sigma_{yy}}{R_2} \right).
\label{eq:contact_force}
\end{equation}
Thus the integrated contact force $F_c$ required to vanish over the wafer--substrate interface is 
\begin{equation}
F_c = -d \int_\Omega  d\Omega \ \left(\frac{\sigma_{xx}}{R_1} + \frac{\sigma_{yy}}{R_2} \right) = 0.
\label{eq:integrated_contact_force}
\end{equation}

Equations~\eqref{eq:chi_zeta_relation_aniso_toroidal} and \eqref{eq:integrated_contact_force} can be imposed to the energy minimization by defining a new functional $\mathcal{L} = \mathcal{F} + \lambda_1 f_{\mathrm{c}} + \lambda_2 F_{\mathrm{c}}$ where $\lambda_{1,2} \in \mathds{R}$ are the Lagrange multipliers, $F_c$ is given by Eq.~\eqref{eq:integrated_contact_force} and the constraint 
\begin{equation}
f_{\mathrm{c}} = \mathcal{D}^4\chi + \frac{1}{R_1 R_2} = 0.\label{eq:constraint}
\end{equation}
The stretching energy thus minimized by finding the set of values $\{C_{k}, \lambda_1, \lambda_2 \}$ that solve the linear system
\begin{equation}
\begin{dcases}
\frac{\partial \mathcal{L}}{\partial C_k} = 0 
\\
\frac{\partial \mathcal{L}}{\partial \lambda_{1,2}} = 0
\end{dcases}\label{eq:linear_system}
\end{equation}
thus determining $\chi$ which further fully determines the stress and strain fields \emph{via} Eqs.~\eqref{eq:Hookes_law} and \eqref{eq:stresses} needed for the X-ray diffraction calculations as detailed in Section~\ref{sec:diffraction_computation}.

\subsection{Calculation of the X-ray diffraction curves\label{sec:diffraction_computation}}

In conjunction with the pure bending strain field, the transverse stretching part has a significant contribution to the X-ray diffraction properties of the crystal due to the reactive strain perpendicular to the diffractive crystal planes mediated by the off-diagonal elements of the compliance matrix. According to Hooke's law [Eq.~\eqref{eq:Hookes_law}], these components in terms of the transverse stretching stress are 
\begin{align}
u_{xz} &= \frac{1}{2}\left(S_{41}\sigma_{xx} + S_{42} \sigma_{yy} + S_{46} \sigma_{xy} \right) \label{eq:aniso_uxz}\\
u_{yz} &= \frac{1}{2}\left(S_{51}\sigma_{xx} + S_{52} \sigma_{yy} + S_{56} \sigma_{xy} \right) \label{eq:aniso_uyz} \\
u_{zz} &= S_{31} \sigma_{xx} + S_{32} \sigma_{yy} + S_{36} \sigma_{xy}. \label{eq:aniso_uzz} 
\end{align}
For the isotropic case, the components $u_{xz}$ and $u_{yz}$ vanish and the remaining one reduces to
\begin{equation}\label{eq:u_zz}
u_{zz} = -\frac{\nu(u_{xx}+u_{yy})}{1-\nu} = -\frac{\nu}{E}(\sigma_{xx}+\sigma_{yy})
\end{equation}

In principle, the calculated total strain field of the pure bending and stretching components can be directly used as a deformation term in the Takagi-Taupin equations but it is computationally a daunting task for a three-dimensional macroscopic crystal. However, as shown previously in \cite{Honkanen_2016}, the problem can be reduced into the convolution of the depth-dependent Takagi-Taupin curve and the lateral strain contribution, assuming that the latter one varies sufficiently slowly along the beam path. The wavelength $\lambda$ of the reflection is changed due the presence of constant strain by an amount $\Delta \lambda$ according to Eq.~(11) of \cite{Honkanen_2016}:
\begin{equation}
\frac{\Delta \lambda}{\lambda} = \frac{\partial (\mathbf{u}\cdot \hat{\mathbf{h}})}{\partial s_\parallel} + 
\frac{\partial (\mathbf{u}\cdot\hat{\mathbf{h}})}{\partial s_\perp} \cot \theta_B
\label{eq:lambda_shifts_with_s}
\end{equation}
where $s_\parallel$ and $s_\perp$ are directions parallel and perpendicular to the reciprocal lattice vector $\mathbf{h}$ ($\hat{\mathbf{h}} = \mathbf{h}/|\mathbf{h}|$) and $\theta_B$ is the Bragg angle. Assuming that the beam propagates transversally in the positive $x$-direction,  Assuming that the beam propagates transversally in the positive $x$-direction, Eq.~\eqref{eq:lambda_shifts_with_s} can be written in terms of photon energy $\mathcal{E} = hc/\lambda$ as
\begin{align}
\frac{\Delta \mathcal{E}}{\mathcal{E}} &= 
-\frac{\partial u_z}{\partial z} \cos^2 \phi
-\left(\frac{\partial u_x}{\partial z} + \frac{\partial u_z}{\partial x} \right)
\sin \phi \cos \phi
-\frac{\partial u_x}{\partial x} \sin^2 \phi \nonumber \\
&-\left[\frac{\partial u_z}{\partial x} \cos^2 \phi
+\left(\frac{\partial u_x}{\partial x} - \frac{\partial u_z}{\partial z} \right) \sin \phi \cos \phi - \frac{\partial u_x}{\partial z} \sin^2 \phi \right] \cot \theta_B,
\label{eq:energy_shifts_u_deriv}
\end{align}
where the asymmetry angle $\phi$ is measured between $z$-axis and $\mathbf{h}$, clockwise-positive. Since the strain is assumed to be constant in the volume of interest, the components of the displacement vector can be written as
\begin{equation}
u_x = u_x^{(0)} + u_x^{(1)} x + u_x^{(2)} z \qquad
u_z = u_z^{(0)} + u_z^{(1)} x + u_z^{(2)} z 
\end{equation}
where $u_x^{(i)}$ and $u_z^{(i)}$ are constants with respect to $x$ and $z$. Taking the partial derivatives of $u_x$ and $u_z$ and comparing to $u_{ij} = (\partial_i u_j + \partial_j u_i)/2$ (note that the term containing derivatives of $\zeta$ can be omitted as it is of the second order), we find that
\begin{equation}
u_x = u_x^{(0)} + u_{xx} x + u_x^{(2)} z \qquad
u_z = u_z^{(0)} + (2 u_{xz} - u_x^{(2)}) x + u_{zz} z.
\end{equation}
Since the bottom of the wafer is in contact with the substrate, this means that $u_z = 0$ at the wafer-substrate interface for every $x$. Therefore we find that $ u_x^{(2)} = 2 u_{xz}$ and thus the partial derivatives of $u_x$ and $u_z$ are
\begin{equation}
\frac{\partial u_x}{\partial x} = u_{xx} \qquad \frac{\partial u_x}{\partial z} = 2 u_{xz} \qquad \frac{\partial u_z}{\partial x} = 0 \qquad \frac{\partial u_z}{\partial z} = u_{zz}.
\end{equation}
Substituting these into Eq.~\eqref{eq:energy_shifts_u_deriv} thus allows us to write the energy shift in terms of the strain tensor:
\begin{align}
\frac{\Delta \mathcal{E}}{\mathcal{E}} = 
&- u_{zz} \cos^2 \phi - 2 u_{xz} \sin \phi \cos \phi
- u_{xx} \sin^2 \phi \nonumber \\
&+\left[\left(u_{zz} - u_{xx} \right) \sin \phi \cos \phi + 2 u_{x z} \sin^2 \phi \right] \cot \theta_B
\end{align}
which in the symmetric Bragg case simplifies to
\begin{equation}
\frac{\Delta \mathcal{E}}{\mathcal{E}} = - u_{zz}. \label{eq:energy_shifts}
\end{equation}
The diffraction (or resolution) curve of the whole crystal wafer is then obtained by calculating the distribution $\rho_{\Delta \mathcal{E}}$ of energy shifts $\Delta \mathcal{E}$ over the surface and convolving the resulting distribution with the 1D Takagi-Taupin curve solved for the pure bending solution Eq.~\eqref{eq:pure_bending_partial_derivatives}. Formally $\rho_{\Delta \mathcal{E}}(\varepsilon)$ for a particular energy shift $\epsilon$ is obtained by summing all the surface elements $d\Omega$ whose energy shift $\Delta \mathcal{E} = \varepsilon$ \emph{i.e.} 
\begin{equation}
\rho_{\Delta \mathcal{E}}(\varepsilon) \propto \int_{\Omega} d \Omega \ \delta(\Delta \mathcal{E} - \varepsilon) \label{eq:energy_distribution}
\end{equation}
where $\delta$ is the Dirac delta function and $\Delta \mathcal{E} = \Delta \mathcal{E}(x,y)$ is understood to be a function of position. Similarly, for rocking curve measurements with a monochromatic beam, the shifts in the diffraction angle are
\begin{align}
\Delta \theta = 
&- \left( u_{zz} \cos^2 \phi + 2 u_{xz} \sin \phi \cos \phi
+ u_{xx} \sin^2 \phi \right) \tan \theta_B \nonumber \\
&+ \left(u_{zz} - u_{xx} \right) \sin \phi \cos \phi + 2 u_{x z} \sin^2 \phi  
\end{align}
which in the symmetric Bragg case simplifies to
\begin{equation}
\Delta \theta = - u_{zz} \tan \theta_B \label{eq:angle_shifts}.
\end{equation}
Note that Eq.~\eqref{eq:angle_shifts} ceases to be valid near $\theta_B = \pi/2$ since it is based on the first order Taylor expansion. The corresponding distribution as a function of shift angle $\alpha$ is
\begin{equation}
\rho_{\Delta \theta}(\alpha) \propto \int_{\Omega} d \Omega \ \delta(\Delta \theta - \alpha). \label{eq:angle_distribution}
\end{equation}
The contribution of energy or angular shifts to the resolution in the respective scan domains can be estimated by calculating the standard deviation of the appropriate distribution.

Usually changes in both $\mathcal{E}$ and $\tan \theta_B$ are minute during scans which means that they can be considered constants. Thus the distributions of $\Delta E$ and $\Delta \theta$ differ only by a multiplicative factor. Therefore, for the sake of brevity, only the derivation of the $\Delta E$ distributions is presented in the following section.

\section{Important special cases\label{sec:special_solutions}}

In this section we apply the general framework presented in Section~\ref{sec:theory} to derive a few important results that are especially relevant considering current trends in the contemporary instrument design. Transverse stretching strain and stress fields due to toroidal bending are derived for circular and rectangular wafers of elastically anisotropic materials, due to their prevalent use in the crystal analyser. In addition, their isotropic counterparts are derived and analysed separately to obtain simplified models for better understanding of anisotropic models and quick analytical estimation of various diffraction properties.

In derivations special attention is put on the spherical bending for three reasons: 1) most of the current state-of-the-art TBCA:s belong to this subclass, 2) availability of the experimental diffraction curves, and 3) it is less complicated to derive the more general toroidal models through examining the spherical bending. The last point becomes evident when we examine the energy minimization constraints. By denoting $R \equiv R_1 = R_2$, the first constraint [Eq.~\eqref{eq:constraint}] becomes
\begin{equation}
f_{\mathrm{c}} = \mathcal{D}^4\chi + \frac{1}{R^2} = 0,
\end{equation}
from which the toroidal case can be fully recovered if we replace the spherical bending radius with the geometrical mean of the toroidal bending radii \emph{i.e.} $R \rightarrow \sqrt{R_1 R_2}$. Therefore the only real difference between the toroidal and spherical bending may arise from the second, contact force constraint of Eq.~\eqref{eq:integrated_contact_force}. However, it turns out that in the cases examined in the following, a solution obtained from minimizing the energy using only the first constraint fulfils automatically also the second one. Therefore, it is sufficient to find a solution using the spherical case and to show that it leads to a vanishing contact force in the toroidal case.

For the sake of completeness,  in the spherical case the pure bending solution [Eq.~\eqref{eq:pure_bending_partial_derivatives_toroidal}] becomes
\begin{align}
&\frac{\partial \epsilon_x'}{\partial x} = -\frac{z}{R} \qquad \quad
\frac{\partial \epsilon_z'}{\partial x} = \frac{x}{R} 
\qquad \quad
\frac{\partial \epsilon_z'}{\partial z} = \frac{S_{31}'(S_{12}' - S_{22}') + S_{32}'(S_{12}' - S_{11}')}{S_{11}'S_{22}' - S_{12}' S_{12}'}\frac{z}{R}
\nonumber \\
&\frac{\partial \epsilon_x'}{\partial z}
= -\frac{x}{R}
+ \frac{(S_{51}'\cos \alpha - S_{41}'\sin \alpha )(S_{12}' - S_{22}') + 
(S_{52}'\cos \alpha - S_{42}'\sin \alpha )(S_{12}' - S_{11}') }{S_{11}'S_{22}' - S_{12}' S_{12}'}\frac{z}{R}
\label{eq:pure_bending_partial_derivatives}
\end{align}
where from Eq.~\eqref{eq:alpha_angle_toroidal}
\begin{equation}
\alpha = \frac{1}{2} \atan \left[\frac{2S_{62}(S_{12}-S_{11}) + 2S_{61}(S_{12}-S_{22})}{S_{66}(S_{11} - S_{22}) + S_{62}^2 - S_{61}^2} \right]
\end{equation}
or  $0$ if $S_{61}' = S_{62}' = 0$. In the isotropic case these partial derivatives reduce to
\begin{equation}
\frac{\partial \epsilon_x}{\partial x} = -\frac{z}{R} \qquad
\frac{\partial \epsilon_x}{\partial z} = -\frac{x}{R} \qquad
\frac{\partial \epsilon_z}{\partial x} = \frac{x}{R} \qquad
\frac{\partial \epsilon_z}{\partial z} = \frac{2 \nu }{1 - \nu}\frac{z}{R}
\end{equation}
where $\nu$ is the Poisson ratio. We see that apart from $\partial \epsilon_x / \partial z$, which has no impact in the symmetric Bragg diffraction, the pure bending strain field of a small spherically bent crystal can be identically reproduced by the isotropic model, when the effective Poisson ratio
\begin{equation}
\nu'_{\mathrm{1D\ TTE}} = \frac{S_{31}'(S_{12}' -S_{22}') + S_{32}'(S_{12}' -S_{11}')}{S_{31}'(S_{12}' -S_{22}') + S_{32}'(S_{12}' -S_{11}') + 2 S_{11}' S_{22}' -2 S_{12}'S_{12}'}\label{eq:nu_1D_TTE}
\end{equation}
is used. Note that $\nu'_{\mathrm{1D\ TTE}}$ can vary radically from reflection to reflection but is invariant with respect to the in-plane rotation. Values of $\nu'_{\mathrm{1D\ TTE}}$ for selected reflections of Si and Ge are tabulated for convenience in Table~\ref{tbl:Si_Ge_elastic_constants}.

\subsection{Isotropic circular wafer}
Consider a spherically bent, isotropic circular crystal wafer with the diameter $L$ and bending radius $R$. As per to the general approach, we could use a truncated series in terms of $x/R$ and $y/R$ as an ansatz for the sought-after $\chi$. However, since the physical system possesses the perfect radial planar symmetry, we can also find the exact solution to the problem with relative ease.   

The formal solution to spherical constraint Eq.~\eqref{eq:chi_zeta_relation_iso_toroidal} is the sum of the general solution to the homogeneous biharmonic equation $\nabla^4 \chi_0 = 0$ and any special solution to nonhomogeneous equation. In polar coordinates $(r,\phi)$ the general solution to the homogeneous biharmonic equation is known as the Michell solution \cite{michell_1899}. For a radially symmetric problem, the solution is required to be independent of $\phi$ so the Michell solution simplifies to $\chi_{0} = A_0 r^2 + B_0 r^2 \ln r + C_0 \ln r$, where $A_0$, $B_0$ and $C_0$ are coefficients to be determined. A special solution to Eq.~\eqref{eq:chi_zeta_relation_iso_toroidal} is $\chi_1 = - Er^4/64R^2$, which is easy to see by substitution. Thus the complete radially symmetric solution to Eq.~\eqref{eq:chi_zeta_relation_iso_toroidal} is
\begin{equation}
\chi = \chi_0 + \chi_1 = -\frac{E}{64 R^2}r^4 + A_0 r^2 + B_0 r^2 \ln r + C_0 \ln r.
\end{equation}
The coefficients are can now be found by minimizing the stretching energy. However, the task can be further simplified by examining the components of stress. Since $\sigma_{ij}$ are given by the second derivatives of $\chi$, we can set $B_0=C_0=0$; otherwise we would obtain diverging components of the stress tensor at $r=0$ owing to the logarithmic terms in $\chi$. Thus from Eq.~\eqref{eq:stresses}, we obtain
\begin{equation}
\sigma_{xx} =  -\frac{E}{16 R^2}(x^2 + 3 y^2) + 2 A_0, \quad
\sigma_{xy} = \frac{E}{8 R^2}x y, \quad
\sigma_{yy} =  -\frac{E}{16 R^2}(3 x^2 + y^2) + 2 A_0. \label{eq:circular_stresses_A0}
\end{equation}
Considering the constraints of minimization, we note that the spherical bending is already enforced by the chosen form of $\chi$, so we do not have include the constraint  \eqref{eq:constraint} into the linear system \eqref{eq:linear_system} explicitely. Furthermore, we choose to neglect the contact force constraint \eqref{eq:integrated_contact_force} for now, thus reducing the linear system to a single equation: 
\begin{equation}
\frac{\partial \mathcal{F}}{\partial A_0} = 0.
\end{equation}
Substituting $\partial \sigma_{xx}/\partial A_0 = \partial \sigma_{yy}/\partial A_0 = 2$ and $\partial \sigma_{xy}/\partial A_0 = 0$ to Equation~\eqref{eq:F_derivative}, the condition becomes 
\begin{equation}
 \int_\Omega  d\Omega \ \sigma_{xx} + \sigma_{yy}  =  \int_0^{2 \pi} d \phi \int_0^{L/2} dr \ r \left( 4 A_0 -\frac{E}{4R^2} r^2 \right) = 0,
\end{equation}
where the prefactor $2d (1 -\nu)$ has been dropped out. Carrying out the integration, the streching energy is found to be minimized when
\begin{equation}
A_0 = \frac{E L^2}{128 R^2}. \label{eq:A_0}
\end{equation}

Substituting \eqref{eq:A_0} back to \eqref{eq:circular_stresses_A0}, we thus obtain
\begin{equation}
\sigma_{xx} =  \frac{E}{16 R^2}\left( \frac{L^2}{4} - x^2 - 3 y^2 \right), \quad
\sigma_{xy} = \frac{E}{8 R^2}x y, \quad
\sigma_{yy} =  \frac{E}{16 R^2}\left(\frac{L^2}{4} - 3 x^2 - y^2\right). \label{eq:circular_stresses}
\end{equation}
Substituting these into the Equations~\eqref{eq:lateral_strains} and \eqref{eq:u_zz}, we obtain the following non-zero components of the strain tensor:
\begin{gather}
u_{xx} = \frac{1}{16 R^2}\left[(1-\nu)\frac{L^2}{4} - (1-3\nu)x^2 - (3-\nu)y^2 \right] \\
u_{yy} = \frac{1}{16 R^2}\left[(1-\nu)\frac{L^2}{4} - (1-3\nu)y^2 - (3-\nu)x^2 \right] \\
u_{xy} = \frac{1+\nu}{8 R^2}x y \\
u_{zz} = \frac{\nu}{4R^2}\left(x^2 + y^2 -\frac{L^2}{8} \right)
\end{gather}

Now, as per the discussion in the beginning of the current section, we now attempt to generalise the solution to the toroidal bending by a trivial substitution $R \rightarrow \sqrt{R_1 R_2}$. According to Eq.~\eqref{eq:contact_force}, the contact force between the wafer and the substrate per unit area is thus 
\begin{equation}
P = \frac{E d}{16 R_1^2 R_2^2} \left[ 
\left(3 R_1 + R_2 \right) x^2
+ \left(R_1 + 3 R_2 \right) y^2
- \left(R_1 + R_2 \right)\frac{L^2}{4}
\right]. \label{eq:contact_force_circular_iso}
\end{equation}
Integrating $P$ over the surface of the wafer results in zero net force which means that the previously omitted constraint \eqref{eq:integrated_contact_force} is in fact fulfilled by the solution obtained without its explicit inclusion. We therefore conclude that the solution, even though derived for a spherical bending, is valid also for the toroidal case.\footnote{This is despite the fact that we assumed the ansatz of $\chi$ to be circularly symmetric, as the bending radii enter the free energy minimization only through their product.}

The symmetry considered, it is convenient to give the components of the stress tensor in the cylindrical coordinates as well. Since 
the stress and strain tensors are second-rank contravariant tensors, they transform as
\begin{equation}
T'_{ij} = \sum_{k,l} \frac{\partial x_i'}{\partial x_k}\frac{\partial x_j'}{\partial x_l} T_{kl}
\end{equation}
where $T'_{ij}$ are the components in the new coordinate system $\lbrace x_i' \rbrace$ and $T_{kl}$ are the components in the old system $\lbrace x_k \rbrace$. Therefore in cylindrical coordinates\footnote{The angular coordinate $\phi$ is actually handled here as $r \phi$ in order to keep the physical unit of the coordinates and thus the dimensions of the transformed tensor components consistent with the Cartesian representation.}
\begin{align}
T'_{rr} &=   \cos^2 \phi T_{xx}  + 2 \sin \phi \cos \phi  T_{xy} +  \sin^2 \phi T_{yy} \label{eq:polartensor1} \\
T'_{r \phi} &=  -  \sin \phi \cos \phi T_{xx} + (\cos^2 \phi - \sin^2 \phi) T_{xy} + \sin \phi \cos \phi T_{yy} \\
T'_{\phi \phi} &= \sin^2 \phi T_{xx} - 2 \sin \phi \cos \phi  T_{xy} + \cos^2 \phi T_{yy} \\
T'_{r z} &= \cos \phi T_{xz} + \sin \phi T_{yz} \\
T'_{\phi z} &= - \sin \phi T_{xz} + \cos \phi T_{yz} \\
T'_{z z} &= T_{zz}. \label{eq:polartensor6}
\end{align}
Thus we obtain
\begin{equation}
\sigma_{rr} = \frac{E}{16 R^2}\left(\frac{L^2}{4}-r^2\right) \qquad \sigma_{r \phi} = 0 \qquad \sigma_{\phi \phi} = \frac{E}{16 R^2}\left(\frac{L^2}{4}-3 r^2\right). \label{eq:circular_polar_stresses}
\end{equation}
Similarly for the strain tensor we have
\begin{gather}
u_{rr} = \frac{1}{16 R^2}\left[(1-\nu)\frac{L^2}{4} - (1-3 \nu)r^2 \right] \\
u_{\phi \phi} = \frac{1}{16 R^2}\left[(1-\nu)\frac{L^2}{4} - (3 - \nu)r^2 \right] \\
u_{r \phi} = 0 \\
u_{zz} = \frac{\nu}{4R^2}\left(r^2 -\frac{L^2}{8} \right)
\end{gather}
We find that the radial normal stress $\sigma_{rr}$ vanishes at the edge of the wafer, which is again a natural outcome since the edges are not supported laterally. Also the shear components $\sigma_{r\phi}$ and $u_{r\phi}$ are zero everywhere which can be interpreted that the crystal is not twisted about the $z$-axis. However, the most interesting behaviour is expressed by the angular normal stress $\sigma_{\phi \phi}$ which is negative near the edge and changes sign at $r = L/\sqrt{12}$. This is a natural consequence from the geometrical fact that the flat wafer cannot fit on a toroidal surface without deforming transversally. The derived result tells us that the most energy efficient way to achieve it is to compress angularly near the edge but extend at the middle of the wafer. This is in contrast to the previous work where only an angular compression was assumed \cite{honkanen_14}. The discrepancy arises from the fact that the previous approach was based solely on the geometrical considerations of the spherical bending which does not fix the value of the elastic energy of the wafer. The requirement of the energy minimization does not alter the resolution curve drastically but does lead to \emph{e.g.} non-vanishing integrated contact force on the wafer--substrate interface. The derivation presented in this work is theoretically more sound and thus expected to be physically more accurate.

As a curiosity it is interesting to note that the qualitative behaviour of $u_{rr}$ is different for $\nu < 1/3$ and $\nu > 1/3$. Whereas in the former case the radial strain is largest at the centre of the wafer, in the latter it is largest at the edge. 

Using Equation~\eqref{eq:energy_shifts}, we find that the energy shift $\Delta \mathcal{E}$ as a function of surface position is
\begin{equation}
\frac{\Delta \mathcal{E}}{\mathcal{E}} = - \frac{\nu}{4R^2}\left(r^2 -\frac{L^2}{8} \right).
\end{equation} 
The isocurves of the energy shift are circular as one would expect on the basis of the radial symmetry. Substituting the obtained $\Delta \mathcal{E}$ to Eq.~\eqref{eq:energy_distribution} and carrying out the integration, we find the energy shift distribution
\begin{equation}
\rho_{\Delta \mathcal{E}}(\varepsilon) = \begin{dcases}
\mathrm{constant}, & -\tfrac{\nu L^2\mathcal{E}}{32 R^2} \leq \varepsilon \leq \tfrac{\nu L^2 \mathcal{E}}{32 R^2} \\
0 & \mathrm{otherwise}
\end{dcases}\label{eq:isotropic_circular_eshift_distribution}
\end{equation} 
The found uniform distribution can be used to convolve the 1D Takagi-Taupin solution to predict the diffraction curve of an TBCA.

To quickly estimate the effect of transverse strain to the energy resolution, we note that the variance of a uniform distribution with a width of $w$ is $w^2/12$ and thus the standard deviation of the energy shift distribution \eqref{eq:isotropic_circular_eshift_distribution} is
\begin{equation}
\sigma = \frac{\nu L^2 \mathcal{E}}{32\sqrt{3} R^2}.\label{eq:isotropic_circular_eresolution}
\end{equation}
The standard deviation due to transverse strain can be then combined with the standard deviations of other contributions (1D Takagi-Taupin, incident bandwidth, etc.) by quadratic summing in accordance with the central limit theorem. Usually the full-width-at-half-maximum (FWHM) is used instead of the standard deviation, in the case of which $\sigma$ is to be multiplied by $2 \sqrt{2 \ln 2}$. This underestimates the true FWHM of Eq.~\eqref{eq:isotropic_circular_eshift_distribution} approximately by a factor of 0.68 but, regarding the central limit theorem, gives more accurate contribution to the total FWHM.

\subsection{Anisotropic circular wafer}
The solving procedure follows the same steps for elastically anisotropic crystals as for the isotropic case. However, since the anisotropy of the crystal does not generally follow the symmetry of the wafer, we should relax the circular symmetry requirement for the ansatz of $\chi$ as well. In general, the candidate solution can be written as a polynomial series of $x/R$ and $y/R$:
\begin{equation}\label{eq:chi_series}
\chi(x,y) = \sum_{m,n=0}^\infty C_{m,n} \left(\frac{x}{R}\right)^{m} \left(\frac{y}{R}\right)^{n}
\end{equation}
For a typical crystal analyser $x/R$ and $y/R$ are order of $0.1$ or less. Thus we may opt to truncate the series representation of $\chi$ up to the few lowest orders. Substituting Eq.~\eqref{eq:chi_series} into the nonhomogeneous biharmonic equation~\eqref{eq:chi_zeta_relation_iso_toroidal}, we find that the simplest solution is of the fourth order. Expanded, the ansatz is then
\begin{align}
\chi &= C_{11}xy + 
\frac{1}{2}\left(C_{20} x^2 +C_{02} y^2 + C_{21} x^2y +C_{12} xy^2  +C_{22}x^2 y^2\right) \nonumber \\
 &+ \frac{1}{3}\left(C_{31} x^3 y + C_{13}x y^3 \right) + \frac{1}{6}\left(C_{30} x^3 +C_{03} y^3\right) 
 + \frac{1}{12}\left(C_{40} x^4 +C_{04} y^4 \right) \label{eq:anisotropic_ansatz}
\end{align}
where the numerical prefactors are added for the subsequent convenience. Coefficients $C_{00}$, $C_{10}$, and $C_{01}$ are set to zero since they do not affect the stress tensor components. Using Eq.~\eqref{eq:stresses}, the transverse stress tensor components are
\begin{align}
\sigma_{xx} &= C_{02} + C_{12} x  + C_{22} x^2  +  C_{03} y + 2 C_{13} x y +  C_{04} y^2 \label{eq:sigma_xx_aniso_circular} \\
\sigma_{yy} &= C_{20} + C_{21} y  + C_{22} y^2  +  C_{30} x + 2 C_{31} x y +  C_{40} x^2 \label{eq:sigma_yy_aniso_circular} \\
\sigma_{xy} &= -C_{11} - C_{21} x - C_{12} y - C_{31} x^2 - C_{13} y^2 - 2 C_{22} xy \label{eq:sigma_xy_aniso_circular}
\end{align}
The spherical bending constraint \eqref{eq:constraint} now becomes
\begin{equation}
f_{\mathrm{c}} = S_{11} C_{04} + S_{22} C_{40} + (2S_{12}+S_{66})C_{22} -2(S_{16} C_{13} + S_{26} C_{31})+ \frac{1}{2R^2} = 0.
\label{eq:constraint_anisotropic_circular}
\end{equation}
Omitting the contact force constraint \eqref{eq:integrated_contact_force} at this stage, the coefficients $C_{ij}$ are solved by minimizing the constrained streching energy which is presented in Appendix~\ref{app:min_F_anisotropic}. The resulting stretching stress tensor components are
\begin{equation}
\sigma_{xx} =  \frac{E'}{16 R^2}\left( \frac{L^2}{4} - x^2 - 3 y^2 \right) \quad 
\sigma_{yy} =  \frac{E'}{16 R^2}\left(\frac{L^2}{4} - 3 x^2 - y^2\right) \quad
\sigma_{xy} = \frac{E'}{8 R^2}x y
 \label{eq:circular_stresses_aniso}
\end{equation}
where 
\begin{equation}
E' = \frac{8}{3(S_{11}+S_{22})+2 S_{12}+S_{66}} \label{eq:effective_E}
\end{equation}
which, in comparison to stresses obtained in the isotropic case [Eq.~\eqref{eq:circular_stresses}], can be interpreted as effective Young's modulus. For isotropic crystal $E' = E$ but in general $E' \neq 1/S_{11}$.

Since the form of the obtained stresses is identical to that of the isotropic case, the immediate implication is that the contact force is equivalent to Eq.~\eqref{eq:contact_force_circular_iso} when effective Young's modulus is used. Therefore the obtained anisotropic solution also is generalisable to the toroidal bending by the trivial substitution $R \rightarrow \sqrt{R_1 R_2}$.

Substituting the obtained stresses to Eqs.~\eqref{eq:aniso_uxx}--\eqref{eq:aniso_uxy} and \eqref{eq:aniso_uxz}--\eqref{eq:aniso_uzz} gives the following strain tensor components:
\begin{align}
u_{xx} &= \frac{E'}{16 R^2} \left[ (S_{11}+S_{12})\frac{L^2}{4} - (S_{11} + 3 S_{12}) x^2  -(3 S_{11} + S_{12}) y^2 + 2 S_{16} xy \right] \\
u_{yy} &= \frac{E'}{16 R^2} \left[ (S_{21}+S_{22})\frac{L^2}{4} - (S_{21} + 3 S_{22}) x^2  -(3 S_{21} + S_{22}) y^2 + 2 S_{26} xy \right] \\
u_{zz} &= \frac{E'}{16 R^2} \left[ (S_{31}+S_{32})\frac{L^2}{4} - (S_{31} + 3 S_{32}) x^2  -(3 S_{31} + S_{32}) y^2 + 2 S_{36} xy \right] \\
u_{xz} &= \frac{E'}{32 R^2} \left[ (S_{41}+S_{42})\frac{L^2}{4} - (S_{41} + 3 S_{42}) x^2  -(3 S_{41} + S_{42}) y^2 + 2 S_{46} xy \right] \\
u_{yz} &= \frac{E'}{32 R^2} \left[ (S_{51}+S_{52})\frac{L^2}{4} - (S_{51} + 3 S_{52}) x^2  -(3 S_{51} + S_{52}) y^2 + 2 S_{56} xy \right] \\
u_{xy} &= \frac{E'}{32 R^2} \left[ (S_{61}+S_{62})\frac{L^2}{4} - (S_{61} + 3 S_{62}) x^2  -(3 S_{61} + S_{62}) y^2 + 2 S_{66} xy \right].
\end{align}
Expressed in polar coordinates, the components of the stress tensor are
\begin{equation}
\sigma_{rr} = \frac{E'}{16 R^2}\left(\frac{L^2}{4}-r^2\right) \qquad \sigma_{r \phi} = 0 \qquad \sigma_{\phi \phi} = \frac{E'}{16 R^2}\left(\frac{L^2}{4}-3 r^2\right) \label{eq:circular_polar_stresses_aniso}
\end{equation}
and the most important strain tensor component\footnote{For the sake of brevity, the other components are not presented here as transforming them using Eqs.~\eqref{eq:polartensor1}--\eqref{eq:polartensor6} is straightforward but the results are lengthy and give little extra value to the discussion of the topic at hand.} from the viewpoint of diffraction calculations is given by 
\begin{align}
u_{zz} &= \frac{E'}{16 R^2} \left[ (S_{31}+S_{32})\frac{L^2}{4} - \left[ 2(S_{31} + S_{32}) + \sqrt{(S_{32} -S_{31})^2 + S_{36}^2} \cos (2 \phi + \beta) \right]r^2 \right] \label{eq:uzz_anisotropic_circular_polar_coordinates}
\end{align}
where $\beta = \atan [S_{36}/(S_{32}-S_{31})]$. 

The symmetric stress tensor is expected to be radially symmetric since transversally anisotropic stress would even itself out, as argued previously in \cite{honkanen_14}. However, the symmetry is broken in the strain tensor due to the anisotropic elastic properties of the crystal. Generally the isocurves of $u_{zz}$ are elliptical whereas for the isotropic case they are circular. The derived expression for $u_{zz}$ is otherwise identical to the previously found result in \cite{honkanen_14} except for the constant term proportional to $L^2$. As discussed in the previous subsection, this is due to the fact that in the previous geometrically based method the total elastic energy was not considered. However, it should be noted that the original approach leads to the same solution if the integrated contact force is required to vanish.

\begin{figure}
\centering
\includegraphics[width=0.7\textwidth]{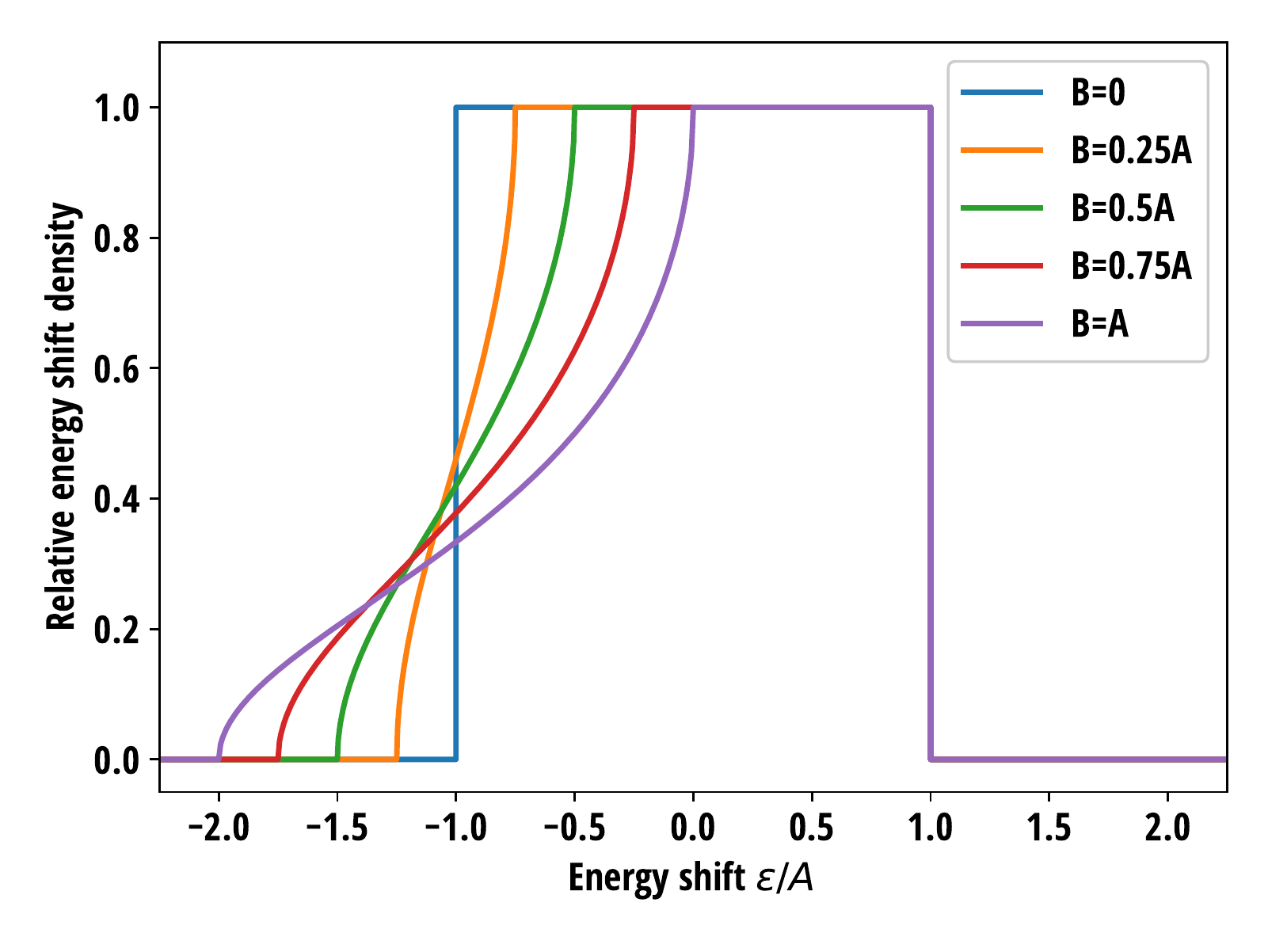}
\caption{Distribution of energy shifts $\rho_{\Delta \mathcal{E}}(\varepsilon)$ for anisotropic circular wafer for various values of $B$.} \label{fig:anisotropic_circular_distribution}
\end{figure}

\begin{figure}
\centering
\includegraphics[width=\textwidth]{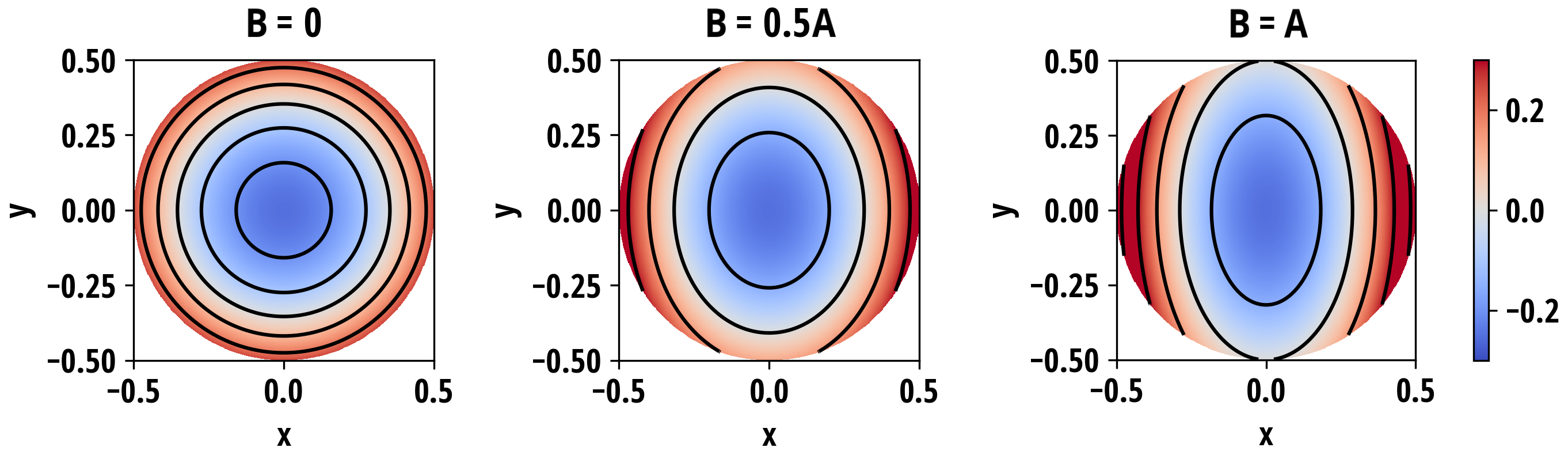}
\caption{Distribution and isocurves of the energy shifts over the anisotropic circular wafer for three different $B/A$ ratios. The gradient of the energy shifts is steepest along the $x$-axis.} \label{fig:uzz_circular_anisotropic}
\end{figure}

As for the isotropic case, the shifts $\Delta \mathcal{E}$ in the diffraction energy are obtained from Eq.~\eqref{eq:energy_shifts}. By substituting to Eq.~\eqref{eq:energy_distribution} and carrying out the radial integration we find that 
\begin{equation}
\rho_{\Delta \mathcal{E}}(\varepsilon) \propto \int_0^{2\pi} d\phi \ \Gamma(\phi,\varepsilon) 
\label{eq:gamma_integral}
\end{equation} 
where
\begin{equation}
\Gamma(\phi,\varepsilon) = \begin{dcases}
\frac{1}{2A + B \cos 2 \phi} &\mathrm{when} \quad -A - B \cos 2 \phi < \varepsilon < A \\
0 & \mathrm{otherwise} 
\end{dcases}
\end{equation}
where the constants are
\begin{equation}
A = -\frac{(S_{31}+S_{32}) E'L^2 \mathcal{E}}{64 R^2} \quad 
B = \frac{E'L^2\mathcal{E}}{64 R^2}\sqrt{(S_{32} -S_{31})^2 + S_{36}^2}.
\end{equation}
Note that $\beta$ has been dropped from the argument of the cosine for simplicity since the integration goes over $2 \pi$. Furthermore from the symmetry of $\cos 2\phi$ it follows that the integrating Eq.~\eqref{eq:gamma_integral} over $2\pi$ is equal to integration over $[0, \pi/2]$ and multiplying the result by 4. Thus
\begin{equation}
\rho_{\Delta \mathcal{E}}(\varepsilon)  \propto  \int_0^{\pi/2} d\phi \ \Gamma(\phi,\varepsilon).\label{eq:gamma_integral2}
\end{equation}
Now since $\acos(x)/2$ can be uniquely mapped over the shortened integration range, 
we can find an angle $0< \phi_0 < \pi/2$ above which the inequality $\varepsilon > -A - B \cos 2 \phi$ ceases to be valid. Therefore we may get rid of the piecewise definition of $\Gamma(\phi,\varepsilon)$ by replacing the upper limit in the integral Equation~\eqref{eq:gamma_integral2} with
\begin{equation}
\phi_0(\varepsilon) = \frac{1}{2} \acos \frac{-A-\varepsilon}{B}
\end{equation}
and thus obtain 
\begin{align}
\rho_{\Delta \mathcal{E}}(\varepsilon) &\propto \int_0^{\phi_0(\varepsilon)} d\phi \ \frac{1}{2A + B \cos 2 \phi} = \frac{1}{\sqrt{4A^2 - B^2}} \atan \left[\frac{(2A-B)\tan \phi_0(\varepsilon)}{\sqrt{4A^2 - B^2}} \right] \nonumber \\
&= \frac{1}{\sqrt{4A^2 - B^2}} \atan \sqrt{\frac{(B-2A)(\varepsilon + A+B)}{(B+2A)(\varepsilon + A -B)}}
\label{eq:gamma_integral3}
\end{align}
when $ -A - B < \varepsilon < - A + B$.
In the interval $-A + B \leq \varepsilon < A$ the integral~\eqref{eq:gamma_integral2} evaluates to a constant which is found by taking the limit $\phi_0(\varepsilon) \rightarrow \pi/2$ of Eq.~\eqref{eq:gamma_integral3}. Thus we find the energy shift distribution
\begin{equation}
\rho_{\Delta \mathcal{E}}(\varepsilon) = k\times\begin{dcases}
\atan \sqrt{\frac{(B-2A)(\varepsilon + A+B)}{(B+2A)(\varepsilon + A -B)}} & -A - B < \varepsilon <  - A + B \\
\frac{\pi}{2}  & -A + B \leq \varepsilon < A \\
0 & \mathrm{otherwise}\label{eq:deltae_shifts_anisotropic_circ}
\end{dcases}
\end{equation}
where $k>0$ is a proportionality constant. Plots of Equation~\eqref{eq:deltae_shifts_anisotropic_circ} with a selected values of $B/A$ are presented in Figure~\ref{fig:anisotropic_circular_distribution}. When $B = 0$, the situation is equivalent to that of the isotropic circular case as the distribution of energy shifts is found to be constant and the energy shift isocurves traced over the crystal surface are perfect circles. For non-zero $B$, the isocurves become elliptical which means that they are intercepted by the circular edge away from the wafer centre, as illustrated in Fig.~\ref{fig:uzz_circular_anisotropic}. The discontinuous isocurves influence the energy shift distribution by introducing a tail on the low energy side of the curve whose prominence is proportional to $B/A$ ratio.

An important practical implication of elliptical isocurves is that there is a specific direction along the surface in which the energy shift varies fastest. Since $S_{31}$ and $S_{32}$ are negative, the gradient of $u_{zz}$ as per to Eq.~\eqref{eq:uzz_anisotropic_circular_polar_coordinates} is steepest in the radial direction when $\cos(2 \phi + \beta) = -1$ \emph{i.e.} $\phi = (-\beta \pm \pi)/2$. This has relevance in regards to the resolution function in cases where the surface area of a TBCA needs to be limited transversally in one direction \emph{e.g.} to minimize the Johann error by masking the surface, or to reduce the space occupied by the analyser by cutting its edges off. To optimize the intrinsic resolution of the analyser, the surface area should be reduced where the gradient is steepest.\footnote{The cut SBCAs in the X-ray Raman scattering spectrometer at the beamline ID20 at ESRF are optimized in this manner \cite{Huotari_2017}.} For example, masking the edges of a spherical Si(660) analyser with 100~mm diameter and 1~m bending radius using a 80~mm wide slit can improve the energy resolution (measured from the standard deviation) by 13\% in near-backscattering conditions if the mask is aligned over the direction of the steepest gradient, which is $[1\overline{1}0]$. However, in the worst-case scenario when the mask is oriented perpendicular to the optimal case, the resolution \emph{degrades} by 3\% in comparison to the unmasked crystal. In the worst case, the resolution of the SBCA in question can thus be 18\% worse than with optimal masking/cutting which is not a negligible detriment. The directions of steepest gradient for selected crystal planes in cubic systems are listed in Table~\ref{tbl:Si_Ge_elastic_constants}.

To estimate the contribution of transverse strain to the energy resolution, the standard deviation of Eq.~\eqref{eq:deltae_shifts_anisotropic_circ} can be calculated from the first and second moments of the normalized distribution, and is found to be
\begin{equation}
\sigma = \frac{\nu' L^2 \mathcal{E}}{32\sqrt{3} R^2}\sqrt{1 + \frac{K^2}{2}} \label{eq:anisotropic_circular_eresolution}
\end{equation}
where we have introduced the effective Poisson's ratio
\begin{equation}
\nu' \equiv - \frac{4(S_{31} + S_{32})}{3(S_{11} + S_{22})+2 S_{12}+S_{66}}\label{eq:effective_nu}
\end{equation}
and the eccentricity factor
\begin{equation}
K \equiv \frac{B}{A} = -\frac{\sqrt{(S_{32}-S_{31})^2 + S_{36}^2}}{S_{31}+S_{32}}.\label{eq:eccentricity_factor}
\end{equation}
The FWHM compliant with the central limit theorem is obtained by multiplying $\sigma$ by $2 \sqrt{2 \ln 2}$. In the isotropic case $\nu' = \nu$ and $K = 0$, thus reducing Eq.~\eqref{eq:anisotropic_circular_eresolution} expectedly to Eq.~\eqref{eq:isotropic_circular_eresolution}. For convenience, Table~\ref{tbl:Si_Ge_elastic_constants} tabulates the effective Young's moduli, Poisson ratios, and eccentricity factors for selected crystal plane directions of Si and Ge.

It should be noted that the effective Poisson ratio $\nu'$ given by Eq.~\eqref{eq:effective_nu} is not identical to $\nu'_{\mathrm{1D\ TTE}}$ used for 1D Takagi-Taupin calculations given by Eq.~\eqref{eq:nu_1D_TTE}. However, the two are well correlated and often very close in value, as can be seen in Table~\ref{tbl:Si_Ge_elastic_constants}.

\begin{figure}
\centering
\includegraphics[width=0.875\textwidth]{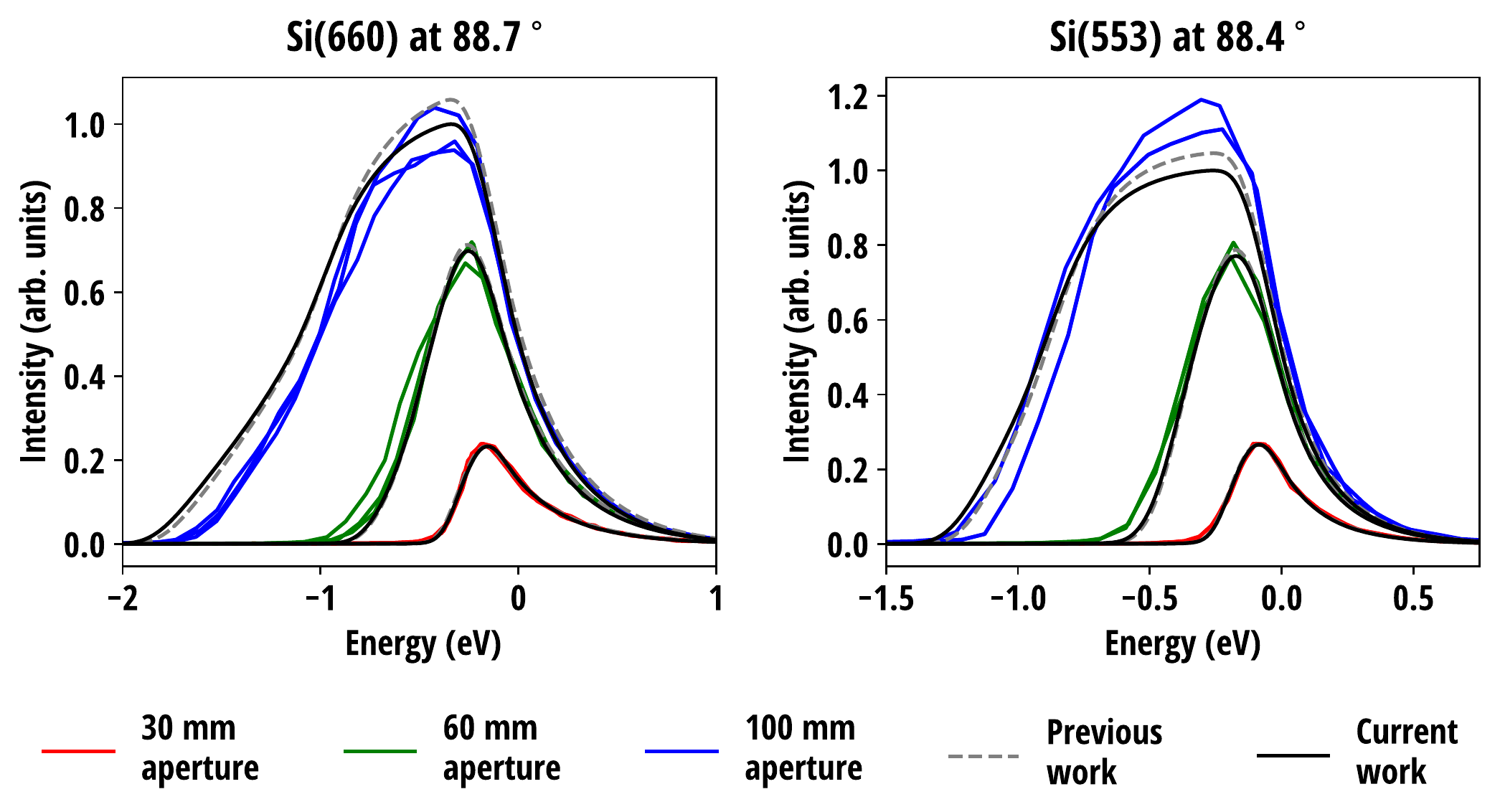}
\vspace{-5mm}
\caption{Measured reflectivity curves of 3 Si(660) and 2 Si(553) SBCAs compared with the predictions of the current and previous work \cite{honkanen_14}. The bending radii were 1~m and the wafer thicknesses were 300~$\upmu$m. The theoretical curves are convolved with the contributions due to the incident bandwidth and Johann error. The centroid energy and the vertical scale of the curves were adjusted as a group to optimize the fit between the theoretical and experimental curves with 30~mm aperture.}
\label{fig:si660_si553}
\end{figure}

\begin{figure}
\centering
\includegraphics[width=0.875\textwidth]{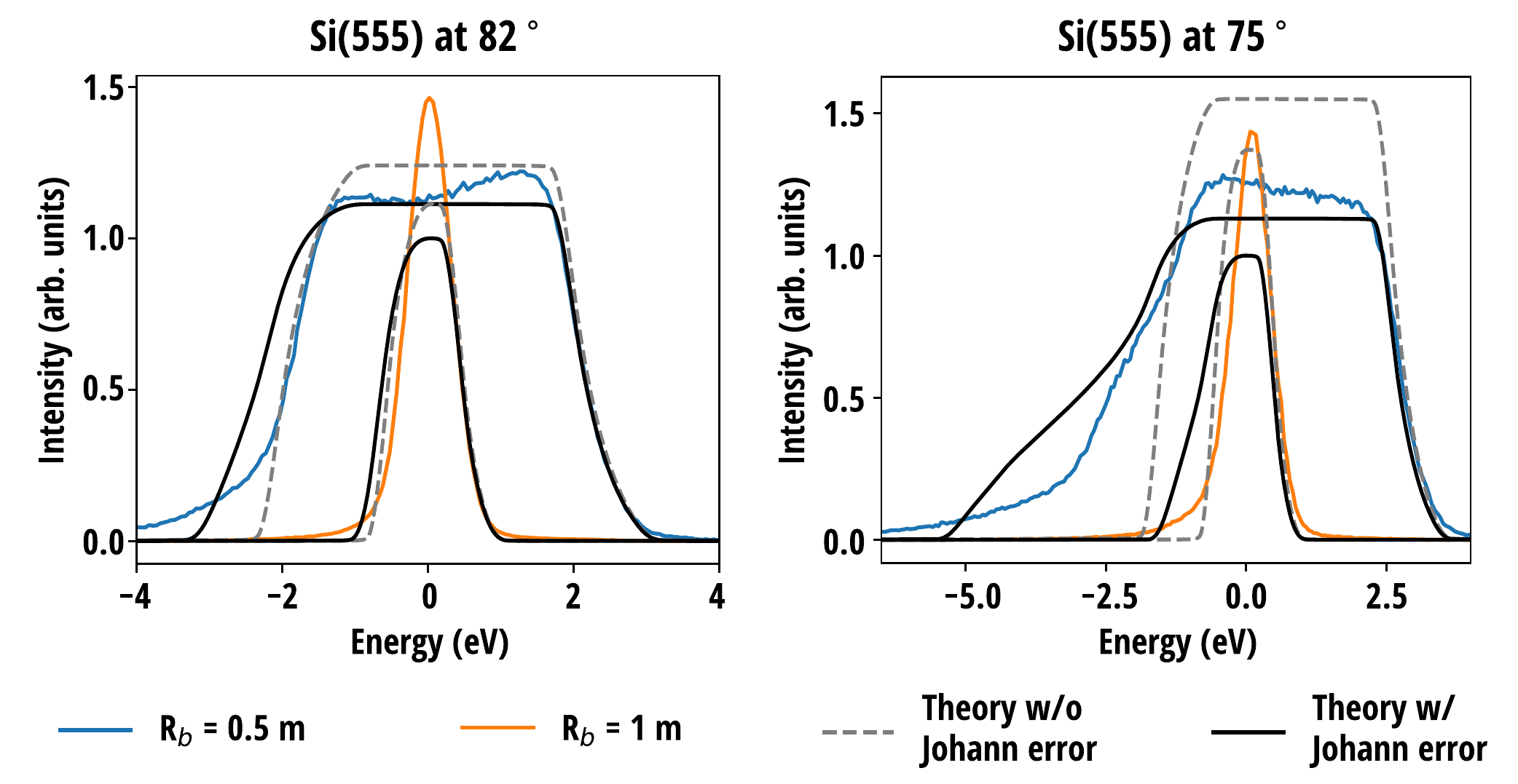}
\vspace{-5mm}
\caption{Calculated reflectivity curves of two circular Si(555) SBCAs with the bending radii of 0.5~m and 1~m at two different Bragg angles in comparison to experimental curves \cite{Rovezzi_2017}. The wafer diameters were 100~mm and the thicknesses 150~$\upmu$m. The centroid energy of the theoretical curves were adjusted separately for 1~m and 0.5~m analysers. The ratio of theoretical integrated intensities of the two SBCAs were scaled according to their solid angle multiplied with their integrated 1D Takagi-Taupin reflectivities.}
\label{fig:si555}
\end{figure}

The predictions of the anisotropic circular model were calculated for four different types of SBCA and compared to two separate experimental data sets acquired at ESRF and first published in \cite{honkanen_14} and \cite{Rovezzi_2017}. In Figure~\ref{fig:si660_si553} are presented the reflectivity curves measured in near-backscattering conditions from three Si(660) and two Si(553) analysers all with the bending radius of 1~m, 100~mm diameter and 300~$\upmu$m wafer thickness. The curves were acquired using two circular masks with aperture diameters of 30~mm and 60~mm, and without mask (aperture 100~mm).  Figure~\ref{fig:si555} presents the comparison of the current model with and without the contribution of Johann error to the reflectivity curves measured at two different Bragg angles of two Si(555) circular analysers with the bending radii of 1~m and 0.5~m. The diameter and thickness of the wafers were 100~mm and 150~$\upmu$m, respectively. Further experimental details are presented in the original sources.

Compared with the previous work which was based on the geometrical considerations and did not account for the minimization of the elastic energy, slight differences between two models are observed but they are found to be less than the variation between different SBCA units, as seen in Fig.~\ref{fig:si660_si553}. This outcludes one explanation put forth in the previous work for the discrepancy between the data and the model at the low-energy tail of the diffraction curve for the full analyser, according to which the observed difference could be due to non-vanishing $\sigma_{rr}$ at the wafer edge in the previous model. One possible explanation to the discrepancy is the imperfections in manufacturing process, as it is found that the figure error in anodically bonded analysers is largest at the edge \cite{Verbeni_2005}. Another explanation could be a slight deviation from the Rowland circle geometry that is not included in the calculations. The latter hypothesis is supported by the data in Fig.~\ref{fig:si555} where the deviations are more prominent. According to the theory, the stresses and strains due to streching are a factor of 4 larger in a wafer that has half the bending radius than in a wafer otherwise identical. Even for considerably higher transverse stress, the theory predicts correctly the observed boxcar shape and its width for the measured 0.5~m Si(555) analyser. The general shape and the width of the predicted 1~m Si(555) curve are in line with the measurements but is not as precise as for the set of Si(660) and Si(553) analysers in Fig.~\ref{fig:si660_si553}. The most probable reason for this is the contribution of aforementioned deviation from the Rowland circle geometry, the effect of which is amplified at lower Bragg angles. In the experimental description, it is mentioned that the radius of the Rowland circle was adjusted by optimizing the product of total counts and peak intensity divided by the FWHM for each analyser \cite{Rovezzi_2017}. Since the different contributions to the energy resolution of an SBCA are not truly independent of each other, such an optimization can lead to partial cancellation of some contribution by another and thus lead to a better resolution than expected in the nominally optimal configuration. Therefore to accurately characterise the elastic contribution to resolution functions of SBCAs, the near-backscattering condition is recommended to minimise the geometrical effects.

\vspace{-5mm}

\begin{table}[]
\begin{sideways}
\parbox{\textheight}{ \centering
\begin{tabular}{ccrrrrrrrr}
   &       & \multicolumn{4}{c}{Si}                                                                                                                  & \multicolumn{4}{c}{Ge}                                                                                                                  \\
$(hkl)$       & $\phi_\mathrm{max}$ & \multicolumn{1}{l}{$\qquad E'$ (GPa)} & \multicolumn{1}{l}{$\nu'$} & \multicolumn{1}{l}{$\nu'_{\mathrm{1D\ TTE}}$} & \multicolumn{1}{l}{$K$} & \multicolumn{1}{l}{$\qquad E'$ (GPa)} & \multicolumn{1}{l}{$\nu'$} & \multicolumn{1}{l}{$\nu'_{\mathrm{1D\ TTE}}$} & \multicolumn{1}{l}{$K$} \\ \hline
{(}100{)} & -- & 147.14                       & 0.3146                                               & 0.2783               & 0                          & 116.84                       & 0.3129                                               & 0.2731             & 0                            \\
{(}110{)} & $[1\overline{1}0]$ & 163.06                       & 0.2043                                               & 0.2032               & 0.7061                     & 131.15                       & 0.1879                                               & 0.1840             & 0.8692                       \\
{(}111{)} & -- & 169.16                       & 0.1621                                               & 0.1801               & 0                          & 136.74                       & 0.1391                                               & 0.1569             & 0                           
\\
{(}210{)} & $[1\overline{2}0]$ & 156.94                       & 0.2467                                               & 0.2372               & 0.3603                     & 125.62                       & 0.2366                                               & 0.2255             & 0.4237
\\
{(}211{)} & $[\overline{1}11]$ & 163.06                       & 0.2043                                               & 0.2107               & 0.2354                     & 131.15                       & 0.1879                                               & 0.1941             & 0.2897
\\
{(}221{)} & $[1\overline{1}0]$ & 166.39                       & 0.1812                                               & 0.1914               & 0.4814                     & 134.20                       & 0.1613                                               & 0.1704             & 0.6139
\\
{(}311{)} & $[\overline{2}33]$ & 156.75                       & 0.2480                                               & 0.2402               & 0.1479                     & 125.44                       & 0.2378                                               & 0.2291             & 0.1737
\\
{(}321{)} & $[\overline{8},11,2]$* & 163.06                       & 0.2043                                               & 0.2086               & 0.4297                     & 131.15                       & 0.1879                                               & 0.1912             & 0.5289
\\
{(}331{)} & $[1\overline{1}0]$ & 164.79                       & 0.1924                                               & 0.1973               & 0.6047                     & 132.73                       & 0.1741                                               & 0.1772             & 0.7572
\\
{(}511{)} & $[\overline{2}55]$ & 151.27                       & 0.2860                                               & 0.2629               & 0.0616                     & 120.52                       & 0.2807                                               & 0.2556             & 0.0704
\\
{(}531{)} & $[\overline{32},51,6]$* & 160.38                       & 0.2229                                               & 0.2210               & 0.4334                     & 128.72                       & 0.2091                                               & 0.2061             & 0.5217
\\
{(}533{)} & $[\overline{6}55]$ & 165.73                       & 0.1859                                               & 0.1974               & 0.2458                     & 133.59                       & 0.1666                                               & 0.1779             & 0.3110
\\
{(}551{)} & $[1\overline{1}0]$ & 163.73                       & 0.1997                                               & 0.2009               & 0.6696                     & 131.77                       & 0.1825                                               & 0.1814             & 0.8294
\\
{(}553{)} & $[1\overline{1}0]$ & 167.32                       & 0.1748                                               & 0.1879               & 0.3892                     & 135.05                       & 0.1539                                               & 0.1662             & 0.5022
\\
{(}731{)} & $[\overline{9},20,3]$* & 155.86                       & 0.2542                                               & 0.2431               & 0.2578                     & 124.64                       & 0.2448                                               & 0.2325             & 0.3012
\\
{(}953{)} & $[\overline{20},31,9]$* & 161.33                       & 0.2163                                               & 0.2181               & 0.3211                     & 129.58                       & 0.2016                                               & 0.2028             & 0.3893
\end{tabular}}
\end{sideways}
\hspace{1cm}
\begin{sideways}
\parbox{\textheight}{\caption{Derived elastic quantities for selected $(hkl)$ normal to the wafer surface of silicon and germanium. $\phi_{\mathrm{max}}$ is the in-plane direction of steepest gradient of $u_{zz}$ which are valid for all cubic systems (directions marked with an asterisk are approximate integer Miller indices). $E'$, $\nu'$ and $K$ are the effective Young modulus [Eq.~\eqref{eq:effective_E}], effective Poisson ratio [Eq.~\eqref{eq:effective_nu}] and eccentricity factor [Eq.~\eqref{eq:eccentricity_factor}], respectively, for the anisotropic circular wafer. $\nu'_{\mathrm{1D\ TTE}}$ is the effective Poisson ratio for 1D Takagi-Taupin calculation [Eq.~\eqref{eq:nu_1D_TTE}] and is valid for arbitrarily shaped spherically bent wafer. The values of elastic matrix elements for Si and Ge are according to \cite{crc_handbook_82nd}. \label{tbl:Si_Ge_elastic_constants}}}
\end{sideways}
\end{table}

\subsection{Isotropic rectangular wafer}

We assume that a spherically bent, rectangular crystal wafer is centred at $x=y=0$ with sides of length $a$ and $b$ aligned parallel with $x$- and $y$-axes, respectively. Since the wafer is symmetric under transformations $x \rightarrow -x$ and $y \rightarrow -y$, we immediately conclude that the series Eq.~\eqref{eq:chi_series} can contain only even terms \emph{i.e.} $C_{m,n} = 0$ if either $m$ or $n$ is odd. Thus we arrive at the fourth-order ansatz 
\begin{equation}\label{eq:chi_rect_ansatz}
\chi(x,y) = \frac{1}{2} \left( C_{20} x^2 + C_{02} y^2 + C_{22} x^2 y^2 \right) + \frac{1}{12} \left( C_{40} x^4 + C_{04} y^4 \right),
\end{equation}
with the added numerical prefactors. In addition, we set $C_{00}=0$ since it has no contribution to the sought-after stress tensor. 
Thus using Equations~\eqref{eq:stresses} we obtain from~\eqref{eq:chi_rect_ansatz}
\begin{equation}
\sigma_{xx} =  C_{22} x^2 + C_{04} y^2 + C_{02}, \quad
\sigma_{xy} = - 2 C_{22} x y, \quad
\sigma_{yy} =  C_{22} y^2  + C_{40} x^2 + C_{20}. \label{eq:isotropic_rect_sigmas}
\end{equation}
The coefficients $C_{ij}$ are found by minimizing the streching energy $\mathcal{F}$ under the requirement that $\chi$ fulfils the non-homogeneous biharmonic equation \eqref{eq:chi_zeta_relation_iso_toroidal}. The details of the minimization are presented in Appendix~\ref{app:min_F_rectangular}. As a result, the following streching strain tensor components are found:
\begin{align}
\sigma_{xx} &= \frac{E}{g R^2}\left[\frac{a^2}{12}- x^2 +\left(\frac{1+\nu}{2}+5\frac{ a^2}{b^2} +\frac{1-\nu}{2} \frac{a^4}{b^4}\right)\left(\frac{b^2}{12}-y^2 \right)\right] \label{eq:sigma_xx_isotropic_rect} \\
\sigma_{yy} &= \frac{E}{g R^2}\left[\frac{b^2}{12}- y^2 +\left(\frac{1+\nu}{2} +5\frac{ b^2}{a^2} +\frac{1-\nu}{2} \frac{b^4}{a^4}\right)\left(\frac{a^2}{12}-x^2 \right)\right] \label{eq:sigma_yy_isotropic_rect} \\
\sigma_{xy} &= \frac{2 E}{g R^2}xy, \label{eq:sigma_xy_isotropic_rect}
\end{align}
where
\begin{equation}
g = 8+10 \left(\frac{a^2}{b^2} + \frac{b^2}{a^2}\right) + (1-\nu)\left(\frac{a^2}{b^2} - \frac{b^2}{a^2}\right)^2.
\end{equation}
We now assume that the obtained solution for the stresses is valid also for the general toroidal bending. From Eq.~\eqref{eq:contact_force} we find the contact force per unit area to be
\begin{align}
P = - \frac{Ed}{gR_1^2R_2^2}\Bigg[
&\left( R_1 \left(\frac{1+\nu}{2} +5\frac{ b^2}{a^2} +\frac{1-\nu}{2} \frac{b^4}{a^4}\right) + R_2\right)\left(\frac{a^2}{12}-x^2 \right) \nonumber \\ +
&\left( R_2 \left(\frac{1+\nu}{2}+5\frac{ a^2}{b^2} +\frac{1-\nu}{2} \frac{a^4}{b^4}\right)
+ R_1 \right)\left(\frac{b^2}{12}-y^2 \right) \Bigg]
\end{align}
Integrating $P$ over the analyser surface results in zero net contact force, which indicates that the constrained omitted in the minimization is automatically fulfilled and  the obtained solution is indeed generalisable to the toroidal by a trivial substitution $R \rightarrow \sqrt{R_1 R_2}$.

An interesting observation is that, contrary to the case of circular wafers, at the edges of the wafer the stress tensor elements describing the normal stress perpendicular to the edge do not vanish. One could argue that the order of the ansatz used is not high enough. However, at least up to the eighth-order, it turns out that requiring the solution to simultaneously to fulfil Eq.~\eqref{eq:chi_zeta_relation_iso_toroidal} and lead to vanishing normal stress at the edges is not possible unless the expansion coefficients of $\chi$ higher that the fourth-order are zero. Fixing the normal component of the stress at the edges completely determines the solution in the fourth order that is necessarily less relaxed than the one obtained through the minimization of energy in Appendix~\ref{app:min_F_rectangular}. Further, it turns out that the integrated contact force  [Eq.~\eqref{eq:integrated_contact_force}] of such a solution is non-zero, which is incompatible with the assumption that the wafer is bent and held onto the spherical substrate by the adhesive force between the wafer and substrate alone. Thus it seems that non-zero normal stress at the edges of the wafer is a real physical part of the rectangular model arising from the mechanical contact between a rectangular wafer and the spherical surface and not an artefact due to the low-order polynomial ansatz.

\begin{figure}
\centering
\includegraphics[width=\textwidth]{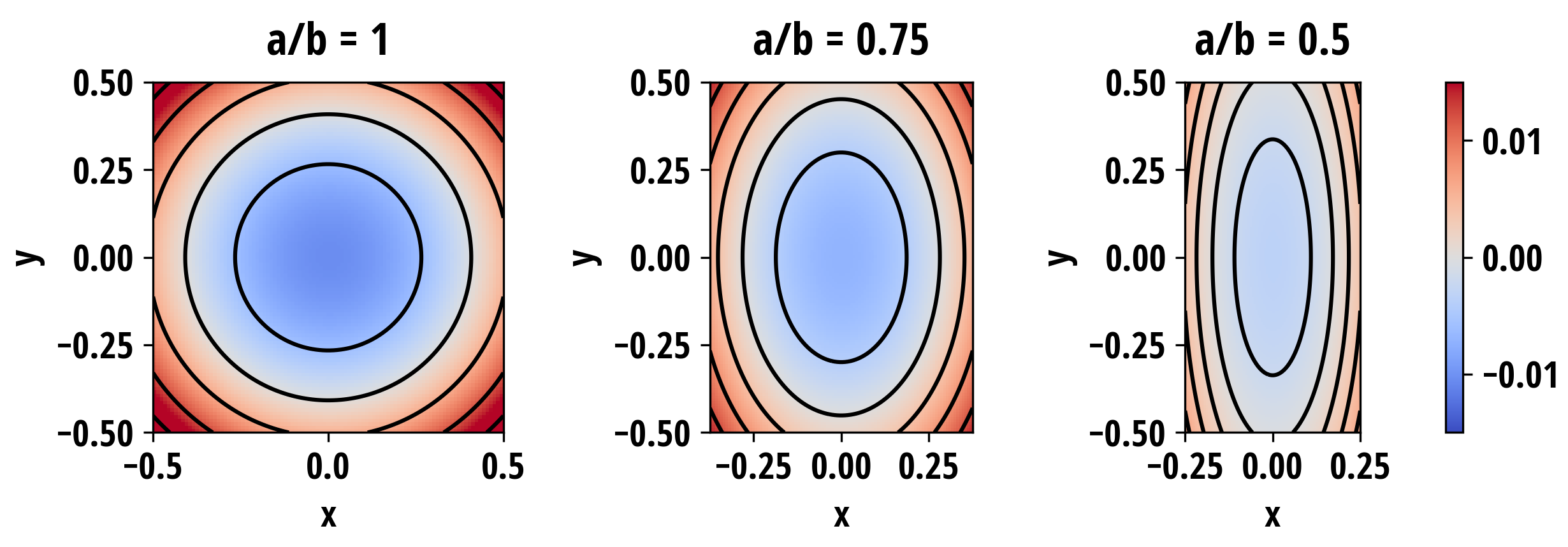}
\caption{The $u_{zz}$ component of the strain for three different wafer side length ratios $a/b$. The Poisson ratio $\nu = 0.25$ was used. Positive (red) values indicate expansion and negative (blue) values indicate the contraction of crystal normal to the surface. Black lines indicate the isocurves of $u_{zz}$.} \label{fig:uzz_rectangular_isotropic}
\end{figure}

Substituting Eqs.~\eqref{eq:sigma_xx_isotropic_rect} and \eqref{eq:sigma_yy_isotropic_rect} to Eq.~\eqref{eq:u_zz}, the most relevant strain tensor component for the diffraction calculations is thus found to be
\begin{align}
u_{zz} = \frac{\nu}{g R^2}\Bigg[&\left(\frac{3+\nu}{2} + 5\frac{b^2}{a^2} +\frac{1-\nu}{2}\frac{b^4}{a^4}  \right)\left( x^2 -\frac{a^2}{12}  \right) \nonumber \\ +
&\left(\frac{3+\nu}{2} + 5\frac{a^2}{b^2} +\frac{1-\nu}{2}\frac{a^4}{b^4} \right)\left( y^2 - \frac{b^2}{12}  \right) \Bigg]\label{eq:uzz_isotropic_rectangular}
\end{align} 
Equation~\eqref{eq:uzz_isotropic_rectangular} for three different $a/b$ ratios is visualised in Figure~\eqref{fig:uzz_rectangular_isotropic}. In general, the crystal planes normal to the surface are compressed in the center of the wafer and expanded at the edges, which is reactionary to transverse extension at the center and contraction at the edges of the wafer via non-zero Poisson's ratio.  The isocurves of $u_{zz}$ are found to be elliptical in shape, albeit being cut near the edges of the wafer. The major axis of the isocurves are along the longer dimension of the wafer and the strain grows fastest along the minor axes. For the special case of $a=b$, the isocurves become circles following the symmetry of the crystal similar to the isotropic circular wafer. It is interesting to note that whereas in the case of circular wafer non-circular isocurves result from the breaking of radial symmetry by the anisotropy of elastic properties of the crystal, for the rectangular wafer it is broken by lifting the 90$^\circ$ rotation symmetry.

\begin{figure}
\centering
\includegraphics[width=0.9\textwidth]{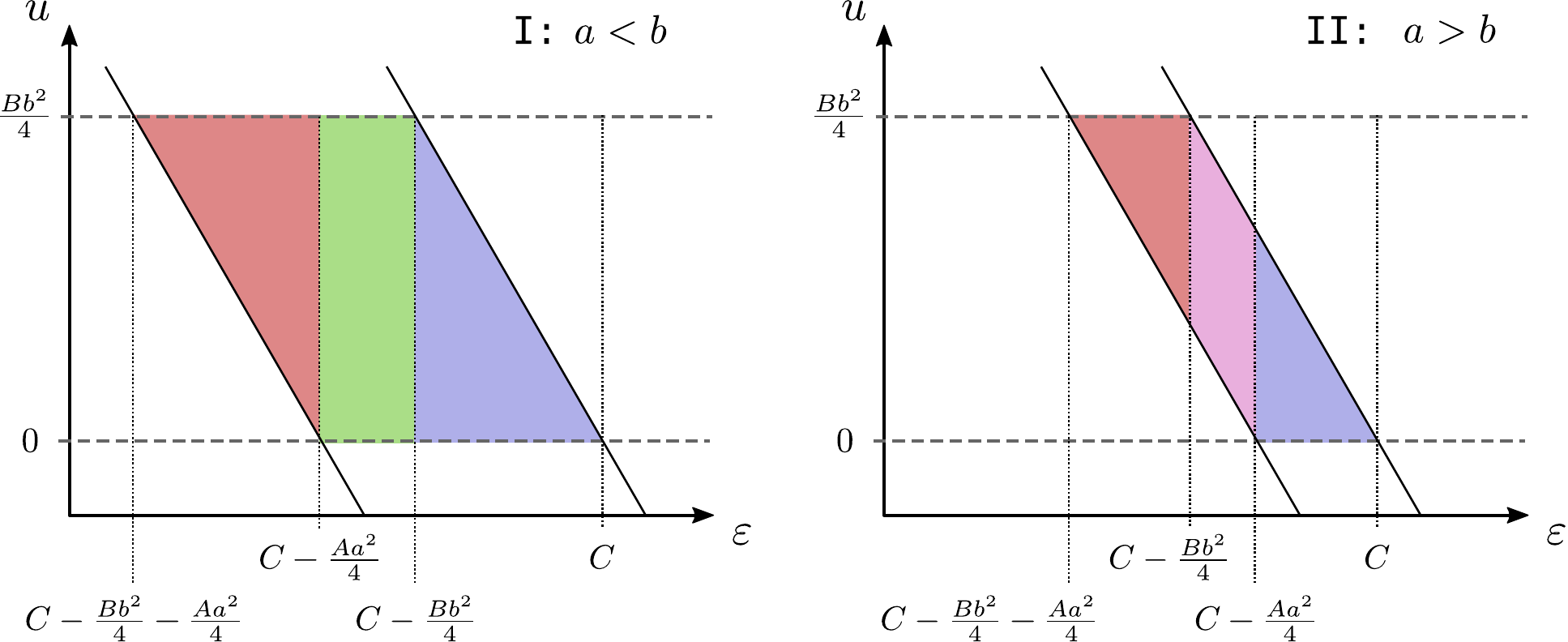}
\caption{Restrictions to the integration range in terms of $u$ imposed by the condition $C - \varepsilon - Aa^2/4 < u < C  - \varepsilon$. The valid integration range presented as colored areas depends linearly on $\epsilon$ in a piecewise manner and is divided into two cases based on whether $A a^2 > Bb^2$ or $A a^2 < Bb^2$. Equivalently, these conditions can be restated in a respective manner as $a<b$ and $a>b$.} \label{fig:isotropic_energy_shift_integration}
\end{figure}

As before, the energy shifts according to Eq.~\eqref{eq:energy_shifts} are $\Delta \mathcal{E} = - u_{zz}\mathcal{E}$. Substituting this to Eq.~\eqref{eq:energy_distribution}, utilizing the symmetries and carrying out the integration along $x$ results to
\begin{equation}
\rho_{\Delta \mathcal{E}}(\varepsilon) \propto 
\int_0^{b/2} dy \ \begin{dcases}
 \frac{1}{\sqrt{C -\varepsilon-By^2}} &\mathrm{when} \quad 0 < C - \varepsilon - B y^2 < \tfrac{Aa^2}{4} \\
0 & \mathrm{otherwise} 
\end{dcases}
\label{eq:gamma_integral_rect}
\end{equation} 
where
\begin{gather}
A = \frac{\nu\mathcal{E}}{gR^2}\left(\frac{3+\nu}{2} + 5 \frac{b^2}{a^2} + \frac{1-\nu}{2} \frac{b^4}{a^4}\right)\quad 
B = \frac{\nu\mathcal{E}}{gR^2}\left(\frac{3+\nu}{2} + 5 \frac{a^2}{b^2} + \frac{1-\nu}{2} \frac{a^4}{b^4}\right)\quad
\nonumber \\ C = \frac{Aa^2 + Bb^2}{12}.
\end{gather}
By performing a change of the integration variable, Eq.~\eqref{eq:gamma_integral_rect} becomes
\begin{equation}
\rho_{\Delta \mathcal{E}}(\varepsilon) \propto 
\int_0^{B b^2/4} du\ \begin{dcases}
 \frac{1}{\sqrt{(C-\varepsilon)u-u^2}} &\mathrm{when} \quad  C - \varepsilon -\tfrac{Aa^2}{4} < u < C  - \varepsilon \\
0 & \mathrm{otherwise} 
\end{dcases}
\label{eq:gamma_integral_rect2}
\end{equation} 
The indefinite solution to the integral is $2 \atan(\sqrt{u/(C - \varepsilon - u)})$ but the integration range is altered by the limits imposed on $u$. Depending whether $Aa^2 > B b^2$ or $Aa^2 < B b^2$, the integration ranges as a piecewise function of $\varepsilon$ can be classified respectively to the Case~$\mathtt{I}$ or $\mathtt{II}$ as indicated by Figure~\ref{fig:isotropic_energy_shift_integration}. It can be shown that $A(a/b)^2 - B$ is a monotonically decreasing function of $a/b$ with the root $a/b = 1$ and thus the conditions simplify to $a<b$ for the Case~$\mathtt{I}$ and $a>b$ for the Case~\texttt{II}. For $a=b$ the cases become identical. As per Fig.~\ref{fig:isotropic_energy_shift_integration}, the integration ranges are 
\begin{align}
\mathrm{Case}\ \mathtt{I}: &\begin{dcases}
 [C-\varepsilon-\tfrac{Aa^2}{4}, \tfrac{Bb^2}{4}] &\mathrm{when} \quad C - \tfrac{Aa^2}{4} - \tfrac{Bb^2}{4} < \varepsilon < C -\tfrac{Aa^2}{4} \\
 [0, \tfrac{Bb^2}{4}] &\mathrm{when} \quad C - \tfrac{Aa^2}{4} \leq \varepsilon \leq C - \tfrac{Bb^2}{4}  \\
 [0, C - \varepsilon] &\mathrm{when} \quad C - \tfrac{Bb^2}{4} < \varepsilon < C
\end{dcases} \\[2ex] 
\mathrm{Case}\ \mathtt{II}: &\begin{dcases}
 [C-\varepsilon-\tfrac{Aa^2}{4}, \tfrac{Bb^2}{4}] &\mathrm{when} \quad C - \tfrac{Aa^2}{4} - \tfrac{Bb^2}{4} < \varepsilon < C -\tfrac{Bb^2}{4} \\
 [C-\varepsilon-\tfrac{Aa^2}{4}, C - \varepsilon] &\mathrm{when} \quad C -\tfrac{Bb^2}{4} \leq \varepsilon \leq C - \tfrac{Aa^2}{4}  \\
 [0, C - \varepsilon] &\mathrm{when} \quad C - \tfrac{Aa^2}{4} < \varepsilon < C
\end{dcases}
\end{align}
Thus the energy shift distribution in the Case~\texttt{I} ($a<b$) is found to be
\begin{equation}
\rho_{\Delta \mathcal{E}}(\varepsilon) = k \times \begin{dcases}
 \frac{\pi}{2} - \atan \sqrt{\tfrac{4(C- \varepsilon)}{B b^2} - 1} - \atan \sqrt{\tfrac{4(C- \varepsilon)}{A a^2} - 1} &\mathrm{when} \ - \tfrac{Aa^2+Bb^2}{6} < \varepsilon < -\tfrac{2 Aa^2 -Bb^2}{12} \\
  \frac{\pi}{2} - \atan \sqrt{\tfrac{4(C- \varepsilon)}{B b^2} - 1}  &\mathrm{when} \ -\tfrac{2 Aa^2 -Bb^2}{12} \leq \varepsilon \leq \tfrac{Aa^2-2Bb^2}{12} \\
   \frac{\pi}{2} &\mathrm{when} \ \tfrac{Aa^2-2Bb^2}{12} < \varepsilon < \tfrac{Aa^2 + Bb^2}{12} \\
   0 & \mathrm{otherwise}
\end{dcases}\label{eq:deltae_distribution_rect}
\end{equation}
where $k>0$ is the proportionality constant. The distribution in the Case~\texttt{II} ($a>b$) is identical to Eq.~\eqref{eq:deltae_distribution_rect} provided that all $Aa^2$ are replaced with $Bb^2$ and vice versa. 

\begin{figure}
\centering
\includegraphics[width=0.9\textwidth]{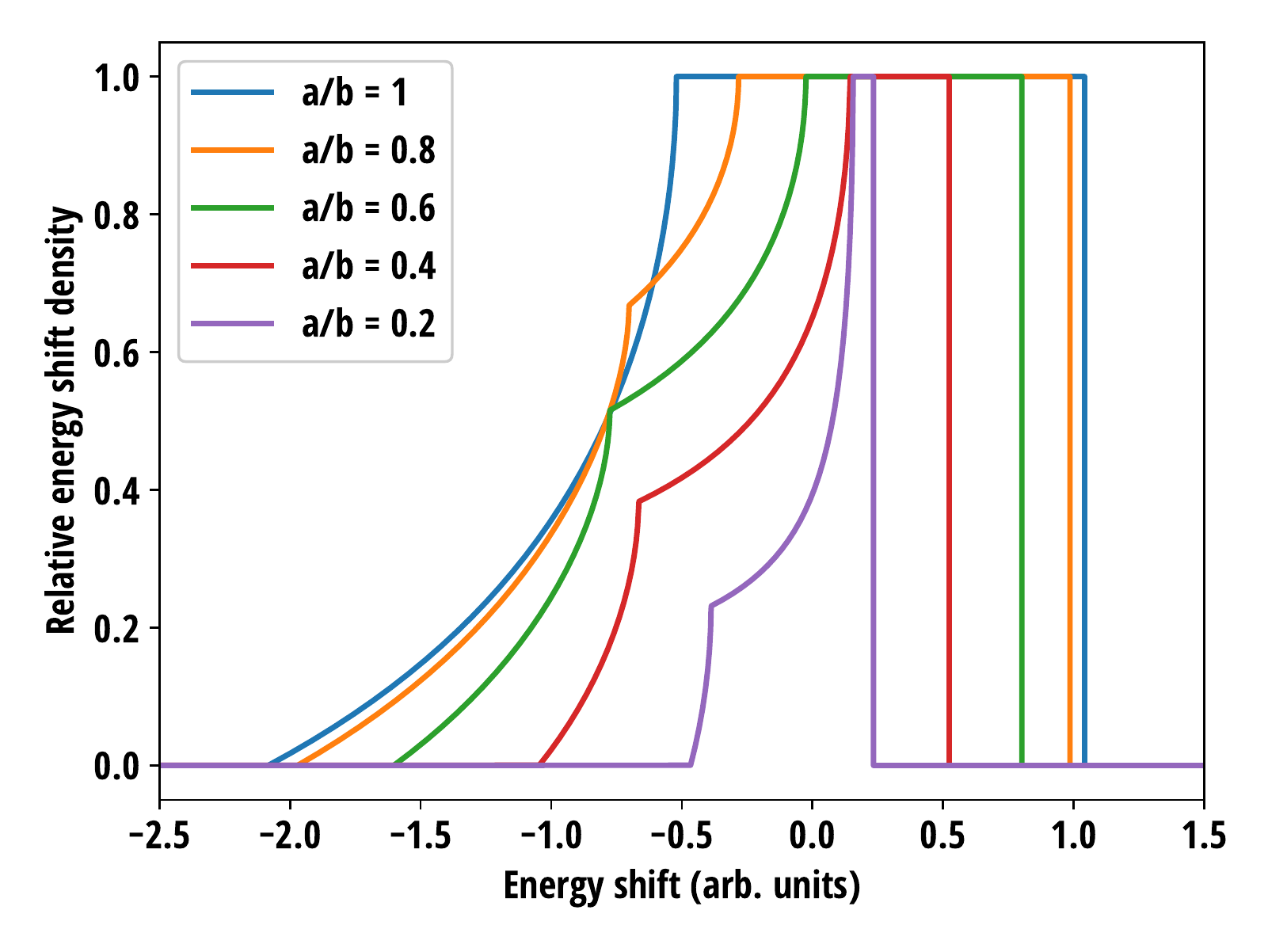}
\caption{The effect of $a/b$ ratio to to the energy shift distribution due to transverse strain in isotropic rectangular crystal. The area of the wafers was kept constant but for visual clarity the curves are normalized to the maximum instead of integrated area. $\nu = 0.25$ was used.} \label{fig:isotropic_rectangular_distribution}
\end{figure}

Examples of energy shift distribution given by Eq.~\eqref{eq:deltae_distribution_rect} are presented in Figure~\ref{fig:isotropic_rectangular_distribution} for rectangular wafers with constant area but various side length ratios. As in the anisotropic circular case, distribution has a flat portion consisting of complete elliptical isocurves and a left-hand side tail caused by the isocurves cropped by the wafer edges (see Fig.~\ref{fig:uzz_rectangular_isotropic}). When $a\neq b$, the tails exhibit a non-differentiable kink due to the isocurves being cropped at different energy shifts along the minor and major axes. Keeping $a/b$ constant, the width of the curve scales proportional to the surface area of the wafer or, equivalently put, to the second power of its linear dimensions and to good accuracy it is directly proportional to the Poisson ratio.

\begin{figure}
\centering
\includegraphics[width=0.9\textwidth]{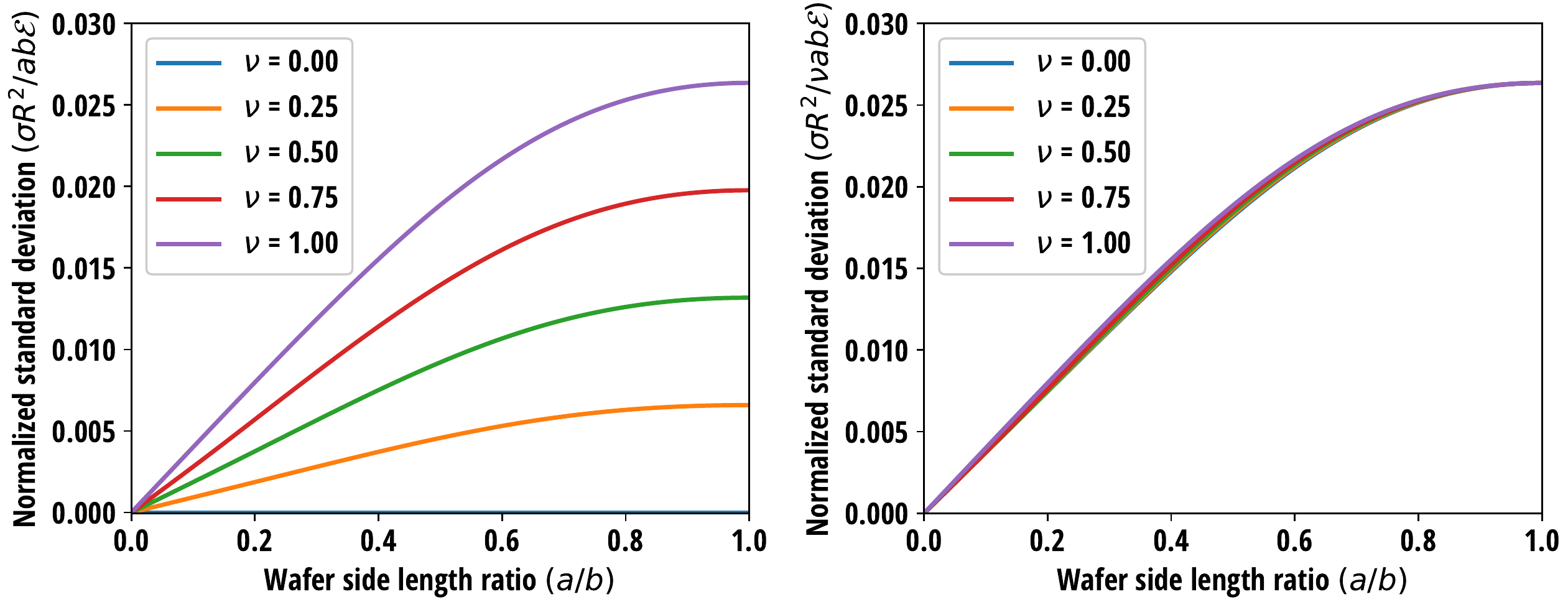}
\caption{\emph{Left:} Normalized standard deviation of the energy shift distribution of isotropic rectangular crystal wafer according to Eq.~\eqref{eq:isotropic_rect_standard_deviation} for various $\nu$. The standard deviation is normalized to the wafer surface area, bending radius and the energy of the incident photons. \emph{Right:} The curves presented on the left panel divided by $\nu$ demonstrating the relative insensitivity of $\sigma$ to the value of $\nu$ apart from scaling.} \label{fig:isotropic_rectangular_resolution}
\end{figure}

The energy resolution of due to transverse streching can be estimated by calculating the standard deviation $\sigma$ of Eq.~\eqref{eq:deltae_distribution_rect}. By integrating the first and second moments of the normalized distribution, we obtain 
\begin{align}
\sigma &= \frac{1}{6\sqrt{5}} \sqrt{A^2 a^4 + B^2 b^4} \nonumber \\
&= \frac{\nu a b \mathcal{E}}{6 g R^2}\sqrt{6+2\nu + \frac{115 + 2\nu -\nu^2}{20} \ehwaz_1 + (1-\nu) \ehwaz_2 + \frac{(1-\nu)^2}{20} \ehwaz_3}
\label{eq:isotropic_rect_standard_deviation}
\end{align}
where
\begin{equation}
\ehwaz_k = \left(\frac{a^2}{b^2}\right)^k + \left(\frac{b^2}{a^2}\right)^k.
\end{equation}
The FWHM compliant with the central limit theorem is obtained by multiplying $\sigma$ by $2 \sqrt{2 \ln 2}$. The standard deviation of the energy shift distribution for various $\nu$ is plotted in the left panel of Fig.~\ref{fig:isotropic_rectangular_resolution} as a function wafer side length ratio. It can be seen that regardless of $\nu$, the standard deviation is maximised and thus the energy resolution of the wafer is the worst when $a/b = 1$ as already indicated by Fig.~\ref{fig:isotropic_rectangular_distribution}. 

Although not obvious from the expression, the square root term divided by $g$ in Eq.~\eqref{eq:isotropic_rect_standard_deviation} is found to depend rather weakly on $\nu$ (Fig.~\ref{fig:isotropic_rectangular_resolution}, right panel). Therefore in practice the exact relation can be approximated to the sufficient extent by the following,  considerably simpler expression
\begin{equation}
\sigma \approx \frac{\nu a b \mathcal{E}}{12\sqrt{2} R^2} \frac{\sqrt{1+0.4\ehwaz_1}}{1+\ehwaz_1}
\label{eq:isotropic_rect_standard_deviation_approx}
\end{equation}
which is accurate within a few precent over the range $0<\nu<1$ being near exact for $\nu=0.5$.

\subsection{Anisotropic rectangular wafer}

In principle the solution for the anisotropic rectangular wafer is obtained by following the same steps as for the anisotropic circular wafer, except for the fact that the integration domain is different. However, it turns out that even though an analytical solution exists, it is too complicated to be practical. Therefore the best approach to anisotropic crystal is to find the solution to the linear system numerically. However, the analytical solution simplifies problem slightly as it turns out that the coefficients $C_{30} = C_{03} = C_{21}= C_{12} = 0$. In addition, the Lagrange multiplier for the integrated contact force $\lambda_2 = 0$ which, in line with the derivations so far, allows us to omit that constraint from the energy minimization.\footnote{An interesting question is whether the integrated contact force vanishes automatically in the minimization of $\mathcal{L} = \mathcal{F} + \lambda_1 f_c$, or does it happen \emph{e.g.} for certain crystal symmetries. Intuitively one could expect the former, as the wafer is easiest to bend by applying a (relatively) weak force normal to the surface but showing this mathematically is out of the scope of this paper.} Thus we can reduce the number of unknowns to be solved from 14 down to 9. We now write the ansatz in the following form
\begin{equation}
\chi = C_{11} x y + \frac{1}{2}\left(C_{20}x^2 + C_{02}y^2 \right)
+ 6 C_{22} x^2 y^2 + 4\left(C_{31}x^3 y + C_{13}x y^3 \right)
+ C_{40} x^4 + C_{04} y^4 \label{eq:chi_ansatz_rectangular_aniso}
\end{equation}
where the numerical prefactors are chosen to simplify the form of the linear system. Substituting the ansatz to Eqs.~\eqref{eq:stresses}, we find the transverse stress tensor components to be
\begin{align}
\sigma_{xx} &= C_{02} + 12 C_{22}x^2 + 24 C_{13} xy + 12 C_{04} y^2 \label{eq:aniso_rect_sigma_xx} \\
\sigma_{yy} &= C_{20} + 12 C_{22}y^2 + 24 C_{31} xy + 12 C_{40} x^2 \\
\sigma_{xy} &= -C_{11}  - 12 C_{31} x^2 - 24 C_{22} xy - 12 C_{13} y^2 \label{eq:aniso_rect_sigma_xy}
\end{align}
The toroidal minimization constraint [Eq.~\eqref{eq:constraint}] is now
\begin{equation}
f_c = 24(2S_{12} + S_{66})C_{22} - 48 S_{26} C_{31} - 48 S_{16} C_{13} +24 S_{22} C_{40} +24 S_{11} C_{04} + \frac{1}{R_1 R_2} = 0 \label{eq:toroidal_constraint_rectangular_aniso}
\end{equation}
The linear system to be minimized is presented in a matrix form Appendix~\ref{app:min_F_rectangular_anisotropic}. After the numerical minimization, the components of the streching tensor are obtained from Eqs.~\eqref{eq:aniso_rect_sigma_xx}--\eqref{eq:aniso_rect_sigma_xy} and the components of the corresponding strain tensor from Eqs.~\eqref{eq:aniso_uxx}--\eqref{eq:aniso_uxy} and \eqref{eq:aniso_uxz}--\eqref{eq:aniso_uzz}. The contact force can be calculated from Eq.~\eqref{eq:contact_force}.

\begin{figure}
\centering
\includegraphics[width=.85\textwidth]{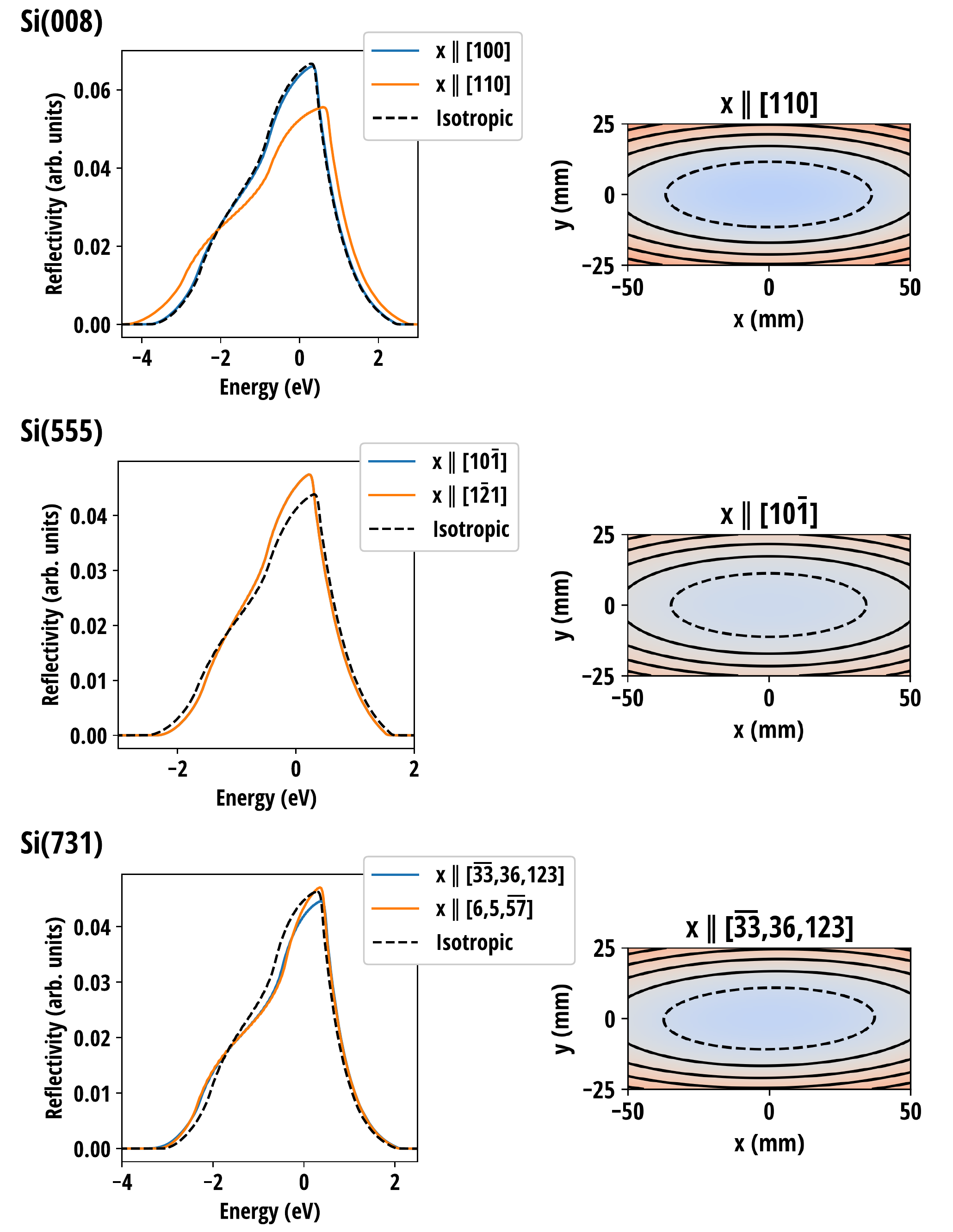}
\caption{\emph{Left column:} Resolution curves of rectangular wafers for three different reflections of Si with selected in-plane crystal orientations aligned with the $x$-axis in comparison to the isotropic model. Note that for Si(555) the curves overlap and the integer indices for Si(731) in-plane directions are approximate. The dimensions of the wafers were set to 100~mm~$\times$~50~mm~$\times$~150~$\upmu$m with the long edges aligned with the $x$-axis. The bending radius was set to 0.5~m and the Bragg angle was 88.5~$^\circ$. The Johann error is omitted. \emph{Right column:} $u_{zz}$-component of the strain tensor over the crystal surface. Red color indicates expansion and blue contraction. Isocurves are marked with solid and dashed black lines.} \label{fig:rectangular_isotropic_anisotropic}
\end{figure}

The predicted reflectivity curves from the anisotropic model are compared to the isotropic one for Si(008), Si(555) and Si(731) reflections in Figure~\ref{fig:rectangular_isotropic_anisotropic}. In general, the isotropic model seems to follow its more intricate anisotropic counterpart rather well when the same Poisson's ratio for the isotropic model is used as for the 1D-Takagi-Taupin curve of the anisotropic model. Unlike for the anisotropic circular crystal, the shape of the resolution curve do not seem to change considerably between different reflections even though their width varies. This is an indication that, as in the isotropic model, the shape of the resolution curve is largely determined by the aspect ratio of the wafer whereas Poisson's ratio scales its width. Furthermore, it seems that the effective Poisson's ratio in the transverse stretching is similar to that of used in 1D-Takagi-Taupin solution, as in the anisotropic circular model.

However, the isotropic model fails to capture some details in the reflectivity curves, most notably the effect of the in-plane orientation of the crystal which for some reflections [\emph{e.g.} Si(008)] can cause a significant effect to the resolution curve of the crystal. Nevertherless, as it is evident from Eqs.~\eqref{eq:aniso_rect_sigma_xx}--\eqref{eq:aniso_rect_sigma_xy}, the isocurves of the transverse stresses, and thus the strains as well, are elliptical in shape as they are in the isotropic case, although for some crystals and orientations the main axes of the ellipses may be inclined with respect to sides of the wafer, as seen for Si(731) in Fig.~\ref{fig:rectangular_isotropic_anisotropic}. 

For the investigated reflections, the isotropic model with the effective 1D-TT Poisson's ratio $\nu'_{\mathrm{1D\ TTE}}$ appears to be a reasonable approximation to the anisotropic one at least for cubic systems. Further theoretical or computational validation is needed to extrapolate the conclusion to other crystal systems.

\subsection{Strip-bent crystal analyser}

\begin{figure}
\centering
\includegraphics[width=0.55\textwidth]{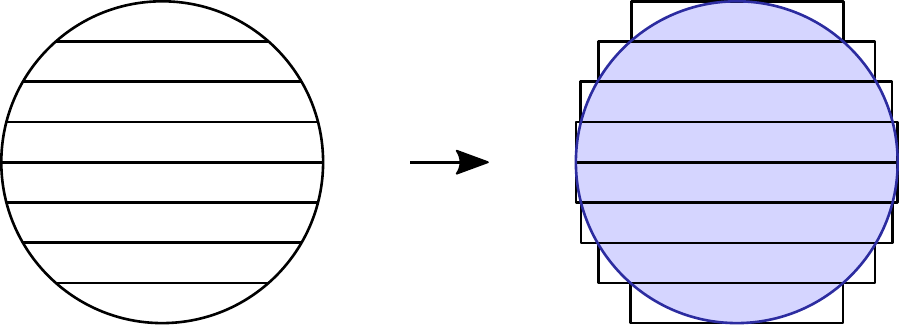}
\caption{Approximation of the strip-bent SBCA using rectangular strips. The wafer is divided into narrow rectangular slices which cover the whole surface area of the circular analyser. The excess parts of the strips are neglected in the approximation.} \label{fig:strip-bent_approximation}
\end{figure}

As seen in Fig.~\eqref{fig:si555}, the transverse stretching can cause a contribution of several eV to the FWHM of the resolution function which is unacceptably large for many spectroscopic purposes. To mitigate the effect of the transverse strain, the surface of the circular wafer can be cut into thin strips before bonding the wafer onto the spherical substrate. The diffraction properties of such a strip-bent analyser can be estimated by approximating the strips by rectangular wafers as presented in Figure~\ref{fig:strip-bent_approximation}. Such an approximation is expected to be most accurate at the center of the analyser where the actual strips are nearly rectangular in shape. The accuracy of the approximation degrades moving laterally perpendicular to the long dimension of the strips but their contribution to the total resolution of the crystal is less significant due to their smaller surface area and thus smaller integrated intensity compared to the medial strips.

There is some freedom in choosing how to approximate the strips with rectangular wafers. Here we have chosen to cover the analyser fully and mask out the parts extending over the circular wafer. This ensures that the approximating strips have the surface area equal to the real strips and allows geometrical errors, such as the Johann error, to be modelled accurately.

\begin{figure}
\centering
\includegraphics[width=\textwidth]{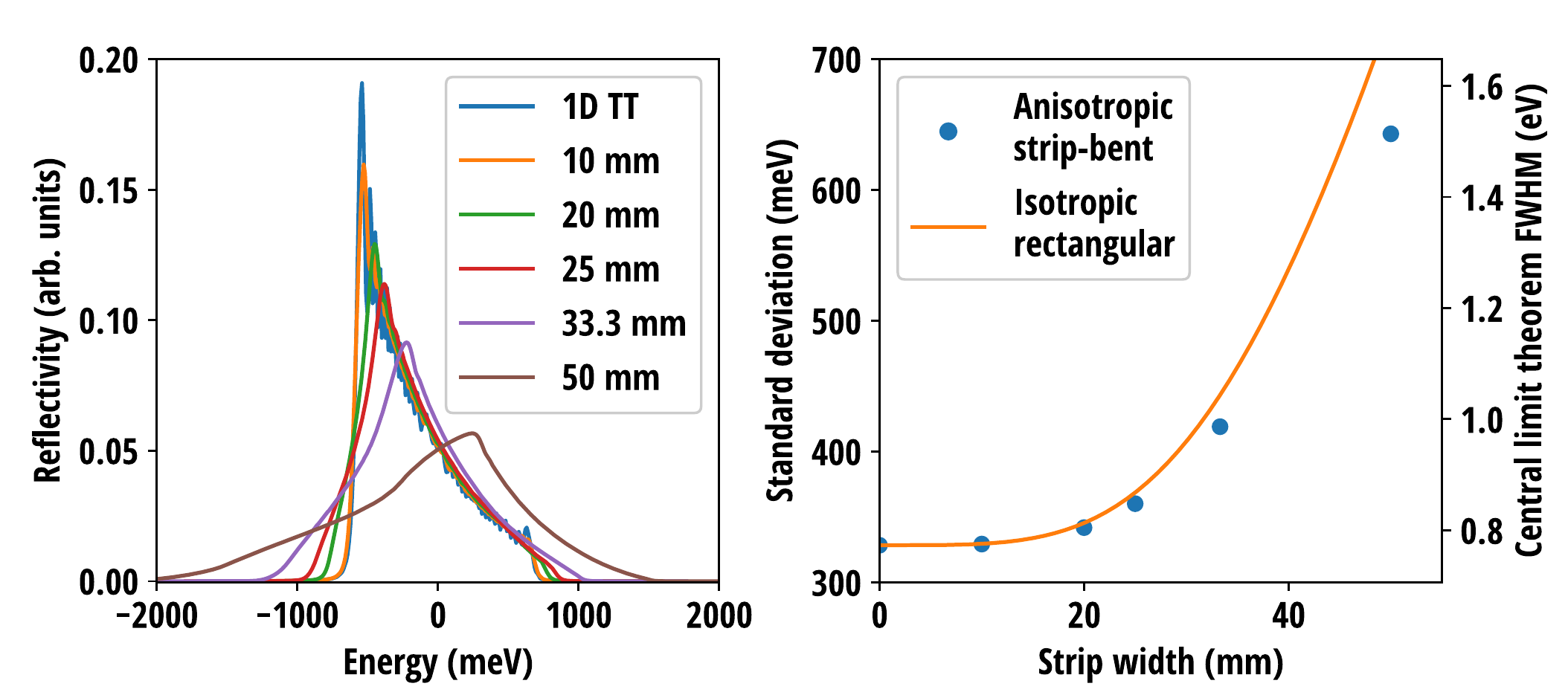}
\caption{\emph{Left panel:} Resolution curves of Si(555) strip-bent analyser with various strip widths compared to the 1D Takagi-Taupin solution. The diameter of the analyser was set to 100~mm, the bending radius to 0.5~m, and the wafer thickness to 150~$\upmu$m. The Bragg angle was chosen to be 88.5$^\circ$ and the Johann error was neglected. \emph{Right panel:} Standard deviations/central limit theorem FWHMs of the resolution curves compared to the prediction based on the isotropic rectangular wafer model with $\nu=0.1801$.} \label{fig:strip_width_comparison}
\end{figure}

In the left panel of Figure~\ref{fig:strip_width_comparison} is presented the calculated resolution curves of strip-bent Si(555) analysers with the bending radius of 0.5~m, diameter of 100~mm and wafer thickness of 150~$\upmu$m at near-backscattering conditions for various strip widths. The strip widths are chosen so that the surface can be divided into an integer number of strips of equal width. As expected, the width of the resolution curve decreases as the strips become narrower and eventually approach the 1D TT-solution calculated with the pure bending deformation. The standard deviations of the resolution curves are presented in the right panel of Fig.~\ref{fig:strip_width_comparison}. Along with the standard deviations is plotted the predicted behaviour according to $\sqrt{\sigma_{\mathrm{1D\ TTE}}^2 + \sigma^2}$ where $\sigma_{\mathrm{1D\ TTE}}$ is the standard deviation of the 1D Takagi-Taupin solution and $\sigma$ is given by the analytical expression Eq.~\eqref{eq:isotropic_rect_standard_deviation_approx} for the isotropic rectangular wafer with the side lengths taken to be strip width and the diameter of the analyser. Poisson's ratio is taken to be the effective Poisson's ratio of the 1D Takagi-Taupin solution. Taking into account that using only the longest strip length overestimates the contribution of shorter strips of the full strip-bent analyser, an accurate correspondence is observed when the strip width is reasonably narrow compared to the analyser width (it is questionable how accurately a masked rectangular wafer estimates the resolution of a hemicircular wafer when the strip width is half the analyser diameter).  

The resolution curves of the state-of-art strip-bent Si(555) analysers manufactured using the anodic bonding techinique were reported in \cite{Rovezzi_2017}. The strip width of the analysers were 15~mm, other physical parameters matching the ones used in the calculations of Fig.~\ref{fig:strip_width_comparison}. Based on the simulations, the transverse stretching begins to contribute notably to the resolution only after the strip width becomes larger than 20~mm, which means that the strip width of the reported analysers is optimal in terms of the stress-relief. The experimental data indeed shows no significant contribution from the transverse strain. From the viewpoint the rectangular wafer and strip-bent model validation, this unfortunately makes a more detailed comparison between the theoretical predictions and the data uninformative.

\section{Reference implementation}

Two open source Python packages, \textsc{pyTTE} and \textsc{tbcalc}, are provided for the low-threshold adoption of the methods to predict the resolution functions of bent isotropic and anisotropic crystal wafers presented in Section~\ref{sec:special_solutions}.  \textsc{pyTTE} calculates 1D X-ray diffraction curves of elastically anisotropic crystals with a depth-depended deformation field in Bragg and Laue geometries by solving the 1D Takagi-Taupin equation using the variable-coeffient ordinary differential equation solver (VODE) with backward differential formula (BDF) method \cite{Brown_1989} as implemented in the \emph{SciPy} library \cite{scipy_ref}. The \textsc{xraylib} library \cite{Schoonjans_2011} is utilized for X-ray diffraction and crystallographic data. \textsc{tbcalc} implements the toroidal bending models to calculate the transverse stress and strain fields and their effect to the resolution curves of isotropic and anisotropic circular and rectangular wafers and strip-bent analysers. The source codes are freely available online at \texttt{https://github.com/aripekka/pyTTE} and \texttt{https://github.com/aripekka/tbcalc}.

\section{Discussion}

Compared to the previous work \cite{honkanen_14, Honkanen_2014b}, the constrained Helmholtz energy minimization approach presented in Section~\ref{sec:theory} offers a straight-forward and general approach to predict the diffraction curves of arbitrarily shaped toroidally bent crystal wafers.
Since toroidal bending encompasses spherical, paraboloidal, and cylindrical bendings, and it can be used as an approximant to many other types of bending as well, the new theory is applicable to the vast majority of crystal optics based on thin, single crystal wafers. In this work we have focused solely on the X-ray diffraction properties but since the Takagi-Taupin theory applies also to neutron diffraction, the method can be extended to neutron optics with minor modifications.

Analytical solutions derived in Section~\ref{sec:special_solutions} give insight into the properties of most commonly encountered circular and rectangular TBCAs and enable both detailed simulations and quick ball-park estimations of the energy resolution. However, the integration domains in the free energy minimization can be easily extended to arbitrarily shaped wafers with numerical methods thus making it possible to simulate even the most unorthodox crystal shapes in search for the optimal instrument performance. 

Nevertheless, even though the method rests on a solid theoretical foundation and is internally consistent, more experimental verification is still needed. Ideally, in order to minimize other effects to the resolution curve, the experiment would be performed in near-backscattering conditions with a $\sigma$-polarized beam and the diffraction curve would be mapped out as a function of position on the crystal surface either using a tightly focused beam or a mask with small aperture in front of the crystal. 

One of the main assumptions in calculating the transverse stretching  is that the wafer is (infinitely) thin and of even thickness everywhere. However, in the practice the wafer is of finite thickness which may vary along the wafer. This variation may be purposeful such as in the case of Johansson type analysers \cite{Johansson_1932,Hosoda_2010}, or inadvertent such as possible imperfections left behind in the manufacturing process. Such variations could be included by replacing the constant thickness $d$ with a function of surface coordinates $d=d(x,y)$ and including it in the integrals of free energy and contact force. Such an approach should work well without further modification if $d(x,y)$ can be written as a low-order polynomial, like in the case of Johansson error, but will require additional additional higher-order terms in the expansion of $\chi$. Alternatively, if the variation in $d(x,y)$ is small, a perturbative approach could turn out to be easier to apply. The latter approach could also be used to include also the figure and slope errors from the perfect toroidal surface due to \emph{e.g.} imperfections in bonding or shape of the substrate \cite{Blasdell_1995, Yumoto_2008, Barrett_2010, Thiess_2010}. More theoretical and computational work is needed to quantify the magnitude of imperfections to the diffraction properties.

\begin{figure}
\centering
\includegraphics[width=0.8\textwidth]{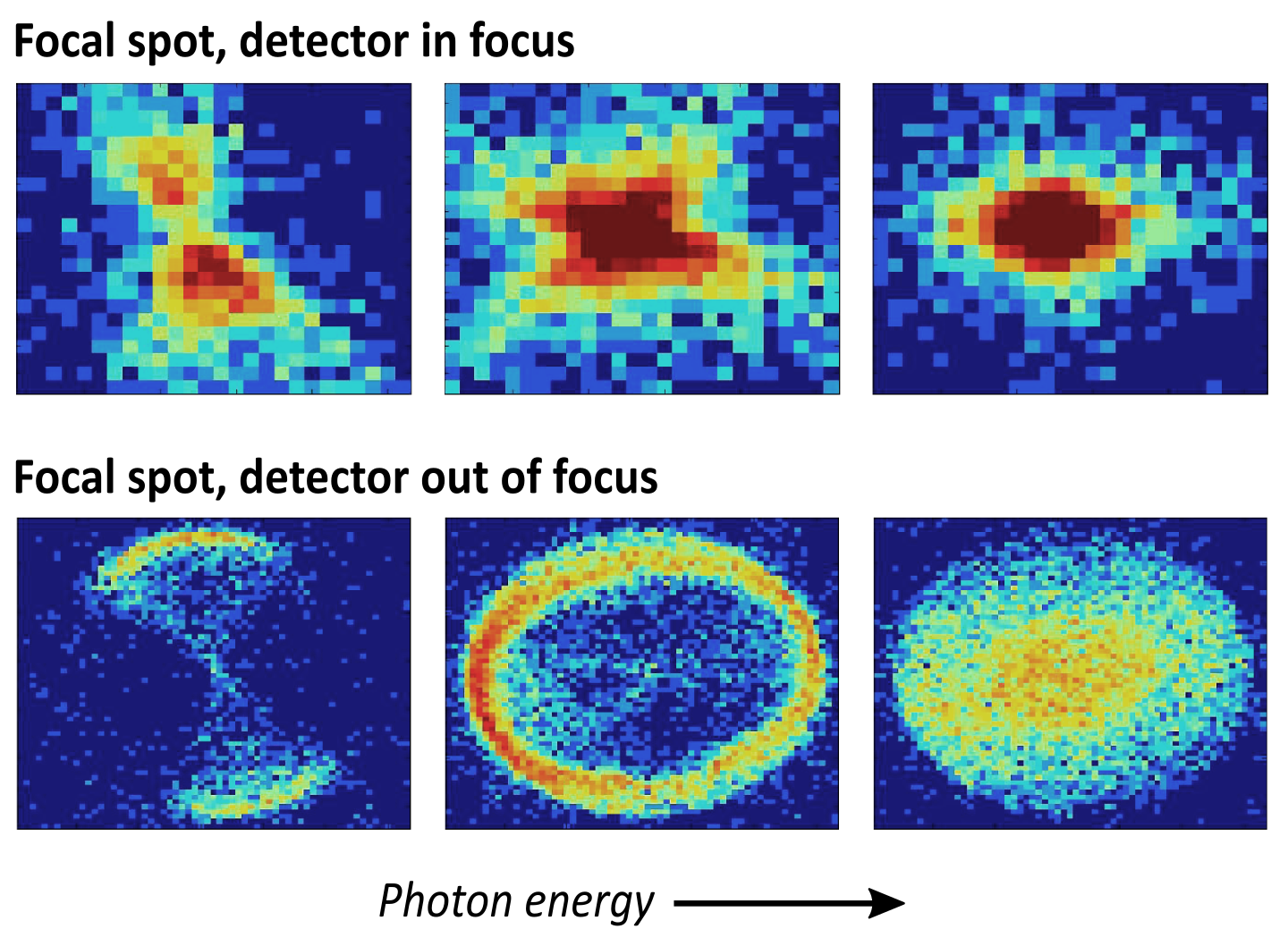}
\caption{A typical focal spot of a circular Si(660) SBCA with bending radius of 1~m and diameter of 100~mm measured in near-backscattering conditions with a position sensitive detector as a function of photon energy. The pixel size is 55~$\upmu$m and the color represents the recorded photon counts in the logarithmic scale. In the top figure the detector was positioned at the focal spot of the detector and in the bottom figure it was moved out of focus, effectively mapping the reflectivity as a function of surface. Note the similarity of the bottom panel with Fig.~\ref{fig:uzz_circular_anisotropic}. The figure is a previously unpublished image from the experimental data set used previously in \cite{honkanen_14} and in Fig.~\ref{fig:si660_si553} of the current work.} \label{fig:focal_spot_hourglass}
\end{figure}

In addition to its energy or angular resolution, another important figure of merit of an crystal analyser is its focusing properties. As presented in Fig.~\ref{fig:focal_spot_hourglass}, when the resolution function of a high quality circular SBCA is measured in the energy domain using a position sensitive detector, one can see the focal spot first appear as a faint hourglass shaped figure at the low energy tail of the resolution curve which then converges into a single spot as the energy is increased.
The orientation of the hourglass pattern corresponds to the direction of steepest gradient of $u_{zz}$ which is a clear indication that transverse stretching can have an effect to the focusing properties of the analyser as well. However, combining the presented method with optical simulations have not been explored in depth for the time being.

\section{Conclusions}

In this work, we have presented a general approach to model the internal strain and stress fields of arbitrarily shaped, toroidally bent crystal wafers and how they can be utilized to predict the diffraction properties of the wafer. Isotropic and anisotropic analytical solutions were derived for circular and rectangular wafers and their properties were discussed in detail focusing on the special case of spherical bending. Comparisons to the available experimental data show that the models can make quantitatively accurate predictions. An open source implementation of the method was discussed and provided.

\ack{\textbf{Acknowledgements}}

APH was funded by the doctoral program in Materials Research and Nanosciences (MATRENA) at the University of Helsinki. The authors want to thank Dr. Mauro Rovezzi\footnote{\includegraphics[scale=0.75]{orcid.png} \url{https://orcid.org/0000-0003-2539-6198}} for providing the Si(555) circular and strip-bent analyser data, and Ari Salop{\"a}{\"a} for providing technical feedback on the Python implementations.

\appendix
\section{Connection of vertical displacement and transverse stress\label{app:lateral_strain}}
From Hooke's law, the transverse components of the strain relate to the stresses by
\begin{align}
u_{xx} &= S_{11} \sigma_{xx} + S_{12} \sigma_{yy} + S_{16} \sigma_{xy} \label{eq:latxx}\\
u_{yy} &= S_{21} \sigma_{xx} + S_{22} \sigma_{yy} + S_{26} \sigma_{xy} \label{eq:latyy}\\
u_{xy} &= \frac{S_{61}}{2} \sigma_{xx} + \frac{S_{62}}{2} \sigma_{yy} + \frac{S_{66}}{2} \sigma_{xy} \label{eq:latxy}
\end{align}
For large deflections, the strain tensor components are
\begin{align}
u_{xx} &= \frac{\partial u_x}{\partial x} + \frac{1}{2}\left(\frac{\partial \zeta}{\partial x} \right)^2 \\
u_{yy} &= \frac{\partial u_y}{\partial y} + \frac{1}{2}\left(\frac{\partial \zeta}{\partial y} \right)^2 \\
u_{xy} &= \frac{1}{2}\left(\frac{\partial u_x}{\partial y} + \frac{\partial u_y}{\partial x} + \frac{\partial \zeta}{\partial x}\frac{\partial \zeta}{\partial y}\right)
\end{align}
Substituting the former and stresses from Eq.~\eqref{eq:stresses} to Eqs.~\eqref{eq:latxx}--\eqref{eq:latxy}
we obtain
\begin{align}
\frac{\partial u_x}{\partial x} + \frac{1}{2}\left(\frac{\partial \zeta}{\partial x} \right)^2 &= S_{11} \frac{\partial^2 \chi}{\partial y^2} + S_{12} \frac{\partial^2 \chi}{\partial x^2} - S_{16} \frac{\partial^2 \chi}{\partial x \partial y} \label{eq:latxx2} \\
\frac{\partial u_y}{\partial y} + \frac{1}{2}\left(\frac{\partial \zeta}{\partial y} \right)^2 &= S_{21} \frac{\partial^2 \chi}{\partial y^2} + S_{22} \frac{\partial^2 \chi}{\partial x^2} - S_{26} \frac{\partial^2 \chi}{\partial x \partial y} \label{eq:latyy2}\\
\frac{\partial u_x}{\partial y} + \frac{\partial u_y}{\partial x} + \frac{\partial \zeta}{\partial x}\frac{\partial \zeta}{\partial y} &= S_{61} \frac{\partial^2 \chi}{\partial y^2} + S_{62} \frac{\partial^2 \chi}{\partial x^2} - S_{66} \frac{\partial^2 \chi}{\partial x \partial y} \label{eq:latxy2}
\end{align}
By taking the partial derivatives $\partial^2/\partial y^2$, $\partial^2/\partial x^2$, and $-\partial^2/\partial x \partial y$ of Eqs.~\eqref{eq:latxx}, \eqref{eq:latyy}, and \eqref{eq:latxy}, respectively, we find
\begin{align}
\frac{\partial^3 u_x}{\partial x \partial y^2} + \frac{\partial \zeta}{\partial x} \frac{\partial^3 \zeta}{\partial x \partial y^2} + \left(\frac{\partial^2 \zeta}{\partial x \partial y} \right)^2 &= S_{11} \frac{\partial^4 \chi}{\partial y^4} + S_{12} \frac{\partial^4 \chi}{\partial x^2 \partial y^2} - S_{16} \frac{\partial^4 \chi}{\partial x \partial y^3} \ \\
\frac{\partial^3 u_y}{\partial x^2 \partial y} + \frac{\partial \zeta}{\partial y} \frac{\partial^3 \zeta}{\partial x^2 \partial y} + \left(\frac{\partial^2 \zeta}{\partial x \partial y} \right)^2 &= S_{21} \frac{\partial^4 \chi}{\partial x^2 \partial y^2} + S_{22} \frac{\partial^4 \chi}{\partial x^4} - S_{26} \frac{\partial^4 \chi}{\partial x^3 \partial y} \\
-\frac{\partial^3 u_x}{\partial x \partial y^2} - \frac{\partial^3 u_y}{\partial x^2 \partial y} - \frac{\partial^3 \zeta}{\partial x^2\partial y}\frac{\partial \zeta}{\partial y}- &\frac{\partial^3 \zeta}{\partial x \partial y^2}\frac{\partial \zeta}{\partial x}
-\frac{\partial^2 \zeta}{\partial x^2}\frac{\partial^2 \zeta}{\partial y^2} -
\left(\frac{\partial^2 \zeta}{\partial x \partial y}\right)^2 = \nonumber \\ - &S_{61} \frac{\partial^4 \chi}{\partial x \partial y^3} - S_{62} \frac{\partial^4 \chi}{\partial x^3 \partial y} + S_{66} \frac{\partial^4 \chi}{\partial x^2 \partial y^2} 
\end{align}
Summing up the equations above sidewise, we thus obtain
\begin{equation}\label{eq:chi_zeta_relation_app}
\mathcal{D}^4\chi = \left(\frac{\partial^2 \zeta}{\partial x \partial y}\right)^2 -\frac{\partial^2 \zeta}{\partial x^2}\frac{\partial^2 \zeta}{\partial y^2},
\end{equation}
where the linear operator $\mathcal{D}^4$ is defined by
\begin{equation}
\mathcal{D}^4 \equiv S_{11} \frac{\partial^4}{\partial y^4} + (2 S_{12} + S_{66}) \frac{\partial^4}{\partial x^2 \partial y^2} +  S_{22} \frac{\partial^4}{\partial x^4}
-2 S_{16}\frac{\partial^4}{\partial x \partial y^3} - 2 S_{26} \frac{\partial^4}{\partial x^3 \partial y}
\end{equation}
and simplified using the symmetry property $S_{ij}=S_{ji}$. Eq.~\eqref{eq:chi_zeta_relation_app} is an anisotropic generalization of Equation~(14.7) in \cite{landau_lifshitz}[p. 53], to which it reduces in the isotropic case.

\section{Contact forces at the wafer--substrate interface\label{app:contact_force}}
Consider a rectangular volume covering the wafer over its whole thickness $d$ in $z$-direction but small in the transverse directions $x$ and $y$. Due to the curved substrate, the surface of the wafer is only approximately aligned with the $xy$-plane and thus the total force acting on the volume element has a small component in $z$ which has to be cancelled by the surface force $P$. 

Let an edge of the volume parallel to $z$ be located at $(x,y)$. Now the normal force acting on the face defined by edges at $(x,y)$ and $(x,y + \Delta y)$, where $\Delta y$ is the side length of the volume in $y$-direction, is
\begin{equation}
F_{x,n} = - d \sin \phi_x \sigma_{xx} \Delta y
\end{equation}
where $\phi_x$ is the inclination of the wafer with respect to the $xy$-plane along $x$. The sign is a result of the outward normal of the face pointing in the negative $x$-direction. Since $\sin \phi_x \approx \partial \zeta / \partial x$, the normal force on the opposite face defined by the 
edges at $(x+\Delta x, y)$ and $(x+\Delta x, y+\Delta y)$, where $\Delta x$ is the side length of the volume in $x$-direction, can be written up to the first order as
\begin{equation}
F'_{x,n} \approx -F_{x,n} + d \frac{\partial^2 \zeta }{\partial x^2} \sigma_{xx}  \Delta y \Delta x
+ d \frac{\partial \zeta }{\partial x} \frac{\partial \sigma_{xx}}{\partial x}  \Delta y \Delta x.
\end{equation}
Performing the same steps for the shear force in the $x$-direction and summing all the forces together, we find the total force due to the stress acting in $x$ is
\begin{equation}
F_x = d \left( \frac{\partial^2 \zeta}{\partial x^2} \sigma_{xx} +
 \frac{\partial \zeta}{\partial x} \frac{\partial \sigma_{xx}}{\partial x} +
 \frac{\partial^2 \zeta}{\partial x \partial y} \sigma_{xy} +
 \frac{\partial \zeta}{\partial x} \frac{\partial \sigma_{xy}}{\partial y}
  \right) \Delta x \Delta y.
\end{equation}
Analogously for the stress acting in the $y$-direction
\begin{equation}
F_y = d \left( \frac{\partial^2 \zeta}{\partial y^2} \sigma_{yy} +
 \frac{\partial \zeta}{\partial y} \frac{\partial \sigma_{yy}}{\partial y} +
 \frac{\partial^2 \zeta}{\partial x \partial y} \sigma_{xy} +
 \frac{\partial \zeta}{\partial y} \frac{\partial \sigma_{xy}}{\partial x}
  \right) \Delta x \Delta y.
\end{equation}
Substituting the Airy stress function $\chi$ from Eq.~\eqref{eq:stresses}, 
we find the total force in the $z$-direction per unit area to be 
\begin{equation}
\frac{F_x + F_y}{\Delta x \Delta y} \approx 
d\left(\frac{\partial^2 \zeta}{\partial x^2} \frac{\partial^2 \chi}{\partial y^2} +  \frac{\partial^2 \zeta}{\partial y^2} \frac{\partial^2 \chi}{\partial x^2}  - 2  \frac{\partial^2 \zeta}{\partial x \partial y} \frac{\partial^2 \chi}{\partial x \partial y} \right)
\end{equation}
which becomes exact at the limit $\Delta x,\Delta y \rightarrow 0$. Substituting the toroidal displacement $\zeta(x,y) = x^2/2R_1 +y^2/2R_2$, we find that the compensating surface force per unit area at the wafer--substrate interface is 
\begin{equation}
P = - d \left( \frac{1}{R_1}\frac{\partial^2 \chi}{\partial y^2} + \frac{1}{R_2}\frac{\partial^2 \chi}{\partial x^2} \right)
= -d \left(\frac{\sigma_{xx}}{R_1} + \frac{\sigma_{yy}}{R_2} \right).\label{eq:surface_forces_toroidal}
\end{equation}
Since thicknesses of the crystal wafers are typically a few hundred micrometers and the bending radii are range from tens to hundreds of centimeters, we may conclude on the basis of the derived expression that the surface forces are indeed negligible compared to the internal stresses.

\section{Minimization of $\mathcal{F}$ for an anisotropic circular wafer\label{app:min_F_anisotropic}}
The streching energy $\mathcal{F}$ is minimized with the toroidal bending constraint $f_c = 0$ by finding the minimum of $\mathcal{L} = \mathcal{F} + \lambda_1 f_c + \lambda_2 F_c$ by solving the linear system given by Eq.~\eqref{eq:linear_system}. It turns out that the contact force constraint $F_c$ can be omitted in the minimization as it is implicitly fulfilled by the solution obtained without it. With the toroidal bending constraint $f_c$ given by Eq.~\eqref{eq:constraint_anisotropic_circular}, the linear system becomes
\begin{align}
&\partial_{11} \mathcal{F} = 0, & &  \partial_{20} \mathcal{F} = \partial_{02} \mathcal{F} = 0, & &\partial_{21} \mathcal{F} = \partial_{12} \mathcal{F} = 0, & &\partial_{30} \mathcal{F} = \partial_{03} \mathcal{F} = 0   \nonumber \\
&\partial_{31} \mathcal{F} - 2 S_{26}\lambda = 0,  &  &\partial_{13} \mathcal{F} - 2 S_{16}\lambda = 0, & & \partial_{40} \mathcal{F} + S_{22} \lambda = 0, & & \partial_{04} \mathcal{F} + S_{11} \lambda = 0 \nonumber \\
&\partial_{22} \mathcal{F} + (2 S_{22} + S_{66})\lambda = 0 
&  &f_c = 0 & &\ & &\
\label{eq:anisotropic_circular_system}
\end{align}
where the shorthand $\partial_k \mathcal{F} \equiv \partial \mathcal{F}/\partial C_{k}$ has been used. By expressing $\sigma_{ij}$ in Eqs.~\eqref{eq:sigma_xx_aniso_circular}--\eqref{eq:sigma_xy_aniso_circular} in polar coordinates, substituting them to Eq.~\eqref{eq:F_derivative_anisotropic}, and carrying out the integration over a circular domain with the diameter $L$, we obtain
\begin{align}
\partial_{11} \mathcal{F} 
= \frac{\pi d L^4}{64}&\Bigg[ -\left(S_{16} + S_{26}  \right) C_{22} - S_{16}C_{04} - S_{26}C_{40} + S_{66}\left(C_{31} + C_{13} \right) \nonumber \\
& -\frac{16}{L^2}\left(S_{16}C_{02}+S_{26}C_{20}-S_{66}C_{11}\right) \Bigg] \\
\partial_{20} \mathcal{F}
= \frac{\pi d L^4}{64}&\Bigg[ \left(S_{12} + S_{22}  \right) C_{22} + S_{12}C_{04} + S_{22}C_{40} - S_{26}\left(C_{31} + C_{13} \right)  \nonumber \\
&+\frac{16}{L^2}\left[S_{12}C_{02}+S_{22}C_{20}-S_{26}C_{11}\right] \Bigg] \\
\partial_{02} \mathcal{F} 
= \frac{\pi d L^4}{64}&\Bigg[ \left(S_{11} + S_{12}  \right) C_{22} + S_{11}C_{04} + S_{12}C_{40} - S_{16}\left(C_{31} + C_{13} \right) \nonumber \\
&+\frac{16}{L^2}\left(S_{11}C_{02}+S_{12}C_{20}-S_{16}C_{11}\right) \Bigg] \\
\partial_{21} \mathcal{F} 
= \frac{\pi d L^4}{64}&\Big[  \left(S_{22} + S_{66} \right)C_{21} -\left(S_{16} + S_{26} \right) C_{12} +S_{12}C_{03} - S_{26}C_{30} \Big]
\\
\partial_{12} \mathcal{F} 
= \frac{\pi d L^4}{64}&\Big[  \left(S_{11} + S_{66} \right)C_{12} -\left(S_{16} + S_{26} \right) C_{21} +S_{12}C_{30} - S_{16}C_{03} \Big]
\\
\partial_{22} \mathcal{F} 
= \frac{\pi d L^4}{64}&\Bigg[ \left(S_{11} + S_{12} \right)C_{02} + \left(S_{12} + S_{22} \right)C_{20} - \left(S_{16}+S_{26}\right)C_{11}
\nonumber \\
&+\frac{L^2}{24} \Big[ \left(3 S_{11} + 2 S_{12} + 3 S_{22} + 4 S_{66} \right) C_{22} - \left(3 S_{16} + 5 S_{26} \right) C_{31} \nonumber \\
&- \left(5 S_{16} + 3 S_{26} \right) C_{13} + \left(3 S_{12} + S_{22} \right) C_{40} + \left(S_{11} + 3 S_{12} \right) C_{04} \Big] \Bigg]
\end{align}
\begin{align}
\partial_{31} \mathcal{F} 
= \frac{\pi d L^4}{64}&\Bigg[ S_{66} C_{11} - S_{16} C_{02} - S_{26} C_{20} -\frac{L^2}{24} \Big[ \left(3 S_{16} + 5 S_{26} \right)C_{22} 
\nonumber \\
&- \left(4 S_{12} + S_{66} \right)C_{13} - \left( 4S_{22} + 3 S_{66} \right) C_{31}  - S_{16}C_{04} -3 S_{26}C_{40}  \Big] \Bigg]
\\
\partial_{13} \mathcal{F} 
= \frac{\pi d L^4}{64}&\Bigg[ S_{66} C_{11} - S_{16} C_{02} - S_{26} C_{20} -\frac{L^2}{24} \Big[ \left(3 S_{26} + 5 S_{16} \right)C_{22} 
\nonumber \\
&- \left(4 S_{12} + S_{66} \right)C_{31} - \left( 4S_{11} + 3 S_{66} \right) C_{13}  - S_{26}C_{04} -3 S_{16}C_{40}  \Big] \Bigg] \\
\partial_{30} \mathcal{F} 
= \frac{\pi d L^4}{64}&\Big[ S_{12}C_{12} - S_{26} C_{21} + S_{22} C_{30} \Big] \\
\partial_{03} \mathcal{F} 
= \frac{\pi d L^4}{64}&\Big[ S_{12}C_{21} - S_{16} C_{12} + S_{11} C_{03} \Big] \\
\partial_{40} \mathcal{F} 
= \frac{\pi d L^4}{64}&\Bigg[ S_{12} C_{02} + S_{22} C_{20} - S_{26} C_{11}  
\nonumber \\
&+\frac{L^2}{24} \Big[ \left(3 S_{12} + S_{22} \right)C_{22} -S_{26}\left(3C_{31} + C_{13} \right) + S_{12} C_{04} + 3 S_{22} C_{40} \Big] \Bigg] \\
\partial_{04} \mathcal{F} 
= \frac{\pi d L^4}{64}&\Bigg[ S_{11} C_{02} + S_{12} C_{20} - S_{16} C_{11}  
\nonumber \\
&+\frac{L^2}{24} \Big[ \left(S_{11} + 3 S_{12} \right)C_{22} -S_{16}\left(3C_{13} + C_{31} \right) + S_{12} C_{40} + 3 S_{11} C_{04} \Big] \Bigg]
\end{align}
Substituting the found derivatives to Eq.~\eqref{eq:anisotropic_circular_system}, the solution to the system is 
\begin{align}
&C_{11} = 0 & &C_{20} = C_{02} = \frac{E' L^2}{64 R^2} & &C_{40} = C_{04} = - \frac{3 E'}{16 R^2} & &C_{22} = -\frac{E'}{16 R^2} \nonumber \\
&C_{30} = C_{03} = 0 & &C_{21} = C_{12} = 0 & &C_{31} = C_{13} = 0 & &\lambda = \frac{\pi d E' L^6}{6144 R^2}
\end{align}
where $R^2 = R_1 R_2$ is the product of bending radii and
\begin{equation}
E' = \frac{8}{3(S_{11}+S_{22})+2 S_{12}+S_{66}}.
\end{equation}

\pagebreak

\section{Minimization of $\mathcal{F}$ for an isotropic rectangular wafer\label{app:min_F_rectangular}}
The streching energy $\mathcal{F}$ is minimized by finding the coefficients $\{C_{ij}, \lambda_1, \lambda_2\}$ which minimize $\mathcal{L} = \mathcal{F} + \lambda_1 f_c + \lambda_2 F_c$ by solving the linear system given by Eq.~\eqref{eq:linear_system}. The constraint $f_c$ is obtained by the requirement that $\chi$ solves Eq.\eqref{eq:chi_zeta_relation_iso_toroidal} \emph{i.e.}
\begin{equation}
f_c = \nabla^4 \chi + \frac{E}{R^2} = 2C_{40} + 4 C_{22} + 2C_{04} + \frac{E}{R^2} = 0,
\end{equation}
where $R^2 = R_1 R_2$ is the product of bending radii. Therefore the equations composing the linear system to be solved are
\begin{align}
&\partial_{20} \mathcal{F} = 0,  & &\partial_{02} \mathcal{F} = 0, & &\partial_{40} \mathcal{F} + 2 \lambda= 0, \nonumber \\
&\partial_{04} \mathcal{F} + 2 \lambda = 0,  & &\partial_{22} \mathcal{F} + 4 \lambda = 0, & &f_c = 0. \label{eq:isotropic_rect_system}
\end{align}
Substituting the stretching stress tensor components given by Eq.~\eqref{eq:isotropic_rect_sigmas}
into the expression of partial derivatives Eq.~\eqref{eq:F_derivative} and carrying out the integration over rectangular domain with linear dimensions $a$ and $b$ in $x$- and $y$-directions, respectively, we thus obtain
\begin{align}
\partial_{20} \mathcal{F} = \frac{abd}{E} &\left[ C_{20} - \nu C_{02} + (C_{40} - \nu C_{22})\frac{a^2}{12} 
+ (C_{22} - \nu C_{04}) \frac{b^2}{12}  \right] \\
\partial_{02} \mathcal{F}  = \frac{abd}{E} &\left[ C_{02} - \nu C_{20} + (C_{22} - \nu C_{40})\frac{a^2}{12} + (C_{04} - \nu C_{22})\frac{b^2}{12} \right] \\
\partial_{04} F = \frac{ab^3d}{12 E} &\left[
C_{02} - \nu C_{20} + (C_{22} - \nu C_{40})\frac{a^2}{12} + 3 (C_{04} - \nu C_{22})\frac{b^2}{20} \right] \\
\partial_{40} F = \frac{a^3 bd}{12 E} &\left[
C_{20} - \nu C_{02} + 3 (C_{40} - \nu C_{22})\frac{a^2}{20} + (C_{22} - \nu C_{04})\frac{b^2}{12} \right] \\
\partial_{22} \mathcal{F} = \frac{abd}{12E} &\Bigg[ 
(C_{02} - \nu C_{20})a^2 + (C_{20} - \nu C_{02})b^2
+ 3(C_{22} - \nu C_{40})\frac{a^4}{20} \nonumber \\
& + \Big[C_{04} + C_{40} + (8  + 6 \nu) C_{22}\Big]\frac{a^2 b^2}{12}  +  3(C_{22} - \nu C_{04})\frac{b^4}{20} \Bigg]
\end{align}
Substituting the calculated derivatives to Eq.~\eqref{eq:isotropic_rect_system}, the solution to the system is
\begin{align}
&C_{20} = \frac{E}{24 g R^2} \left[(1+\nu)a^2 + 12 b^2 +(1-\nu)\frac{b^4}{a^2} \right], \
C_{40} = -\frac{E}{2 g R^2} \left[ 1+\nu + 10\frac{b^2}{a^2} + (1-\nu)\frac{b^4}{a^4}\right], \nonumber \\
&C_{02} = \frac{E}{24 g R^2} \left[(1+\nu)b^2 + 12 a^2 +(1-\nu)\frac{a^4}{b^2} \right], \
C_{04} = -\frac{E}{2 g R^2} \left[ 1+\nu + 10\frac{a^2}{b^2} + (1-\nu)\frac{a^4}{b^4}\right], \nonumber \\
&C_{22} = -\frac{E}{g R^2}, \qquad \lambda = \frac{d}{720 g R^2} \left[(1-\nu)(a^5b+ ab^5) + 10 a^3b^3  \right]
\end{align}
where
\begin{equation}
g = 8+10 \left(\frac{a^2}{b^2} + \frac{b^2}{a^2}\right) + (1-\nu)\left(\frac{a^2}{b^2} - \frac{b^2}{a^2}\right)^2
\end{equation}

\section{Minimization of $\mathcal{F}$ for an anisotropic rectangular wafer\label{app:min_F_rectangular_anisotropic}}
The streching energy $\mathcal{F}$ is minimized with the toroidal bending constraint $f_c = 0$ by finding the minimum of $\mathcal{L} = \mathcal{F} + \lambda_1 f_c$ by solving the linear system given by Eq.~\eqref{eq:linear_system}. 
Using the ansatz from Eq.~\eqref{eq:chi_ansatz_rectangular_aniso} for $\chi$, the constraint from Eq.~\eqref{eq:toroidal_constraint_rectangular_aniso}, and rewriting the Lagrange multiplier $\lambda_1 \rightarrow \lambda_1 abd/120$, we may reformulate the problem as solving the matrix equation $\Lambda  \mathbf{C} = \mathbf{b}$ in terms of $\mathbf{C}$ where
\begin{equation}
\mathbf{C} = \left[ \begin{matrix}
C_{11} & C_{20} & C_{02} & C_{22} & C_{31} & C_{13} & C_{40} & C_{04} & \lambda_1
\end{matrix}
\right]^{\mathrm{T}},
\end{equation}
\begin{equation}
\mathbf{b} = \left[ \begin{matrix}
0 & 0 & 0 & 0 & 0 &0 &0 & 0 & -(24 R_1 R_2)^{-1}
\end{matrix}
\right]^{\mathrm{T}},
\end{equation}
and
\begin{equation}
\Lambda = \left[
\begin{matrix}
 S_{66} & -S_{26} & -S_{16} & \Lambda_{14} & S_{66} a^2 & S_{66} b^2 & -S_{26} a^2 & -S_{16} b^2 & 0 \\
-S_{26} & S_{22} & S_{12} & \Lambda_{24}  & - S_{26} a^2 & -S_{26} b^2 & S_{22} a^2 & S_{12} b^2 & 0 \\
-S_{16} & S_{12} & S_{11} & \Lambda_{34} & - S_{16} a^2 & -S_{16} b^2 & S_{12} a^2 & S_{11} b^2 & 0 \\
 \Lambda_{41} & \Lambda_{42} & \Lambda_{43} & \Lambda_{44} & \Lambda_{45} & \Lambda_{46} & \Lambda_{47} & \Lambda_{48} & \Lambda_{49} \\
5 S_{66} a^2 & -5 S_{26} a^2 & -5 S_{16} a^2 & \Lambda_{54} & \Lambda_{55} & \Lambda_{56} & -9 S_{26} a^4 & -5 S_{16} a^2 b^2 & - 2 S_{26} \\
5 S_{66} b^2 & -5 S_{26} b^2 & -5 S_{16} b^2 & \Lambda_{64} & \Lambda_{65} & \Lambda_{66} & -5 S_{26} a^2 b^2 & -9 S_{16} b^4 & - 2 S_{16} \\
-5 S_{26} a^2 & 5 S_{22} a^2 & 5 S_{12} a^2 & \Lambda_{74} & -9 S_{26} a^4 & -5 S_{26} a^2 b^2 & 9 S_{22} a^4 & 5 S_{12} a^2 b^2 & S_{22} \\
-5 S_{16} b^2 & 5 S_{12} b^2 & 5 S_{11} b^2 & \Lambda_{84} & -5 S_{16} a^2 b^2 & -9 S_{16} b^4 & 5 S_{12} a^2 b^2 & 9 S_{11} b^4 & S_{11} \\
0 & 0 & 0 & \Lambda_{94} & -2 S_{26} & -2 S_{16} & S_{22} & S_{11} & 0
\end{matrix}
\right]
\end{equation}
with
\begin{align}
\Lambda_{14} &= -S_{16} a^2 -S_{26}b^2 & \Lambda_{24} &= S_{12}a^2 + S_{22}b^2 \nonumber  \\
\Lambda_{34} &= S_{11}a^2 + S_{12}b^2 & \Lambda_{41} &= -5 S_{16}a^2 - 5 S_{26}b^2 \nonumber  \\
\Lambda_{42} &= 5 S_{12}a^2 + 5 S_{22}b^2 & \Lambda_{43} &= 5 S_{11}a^2 + 5 S_{12}b^2 \nonumber \\
\Lambda_{44} &= 9 S_{11} a^4 + 9 S_{22} b^4 + 10 (S_{12}+2 S_{66})a^2 b^2 & 
\Lambda_{45} &= -9 S_{16} a^4 -25 S_{26} a^2 b^2 \nonumber  \\
\Lambda_{46} &= -25 S_{16}a^2b^2 -9 S_{26} b^4 & \Lambda_{47} &= 9 S_{12} a^4 + 5 S_{22} a^2 b^2 \nonumber \\
\Lambda_{48} &= 5 S_{11} a^2 b^2 + 9 S_{12} b^4 & \Lambda_{49} &= 2 S_{12} + S_{66} \nonumber \\
\Lambda_{54} &= -9 S_{16} a^4 -25 S_{26} a^2 b^2 & \Lambda_{55} &= 9 S_{66} a^4 + 20 S_{22} a^2 b^2 \nonumber \\
\Lambda_{56} &= 5(4 S_{12} + S_{66} ) a^2 b^2 & \Lambda_{64} &=  -25 S_{16} a^2 b^2 -9 S_{26} b^4 \nonumber  \\
\Lambda_{65} &= 5(4 S_{12} + S_{66}) a^2 b^2 & \Lambda_{66} &= 20 S_{11} a^2 b^2 + 9 S_{66} b^4 \nonumber \\
\Lambda_{74} &= 9 S_{12} a^4 + 5 S_{22} a^2 b^2 & \Lambda_{84} &= 5 S_{11} a^2 b^2 + 9 S_{12} b^4 \nonumber \\
\Lambda_{94} &=  2 S_{12} + S_{66} & & \nonumber 
\end{align}

\section{Johann error}
\begin{figure}
\centering
\includegraphics[width=0.4\textwidth]{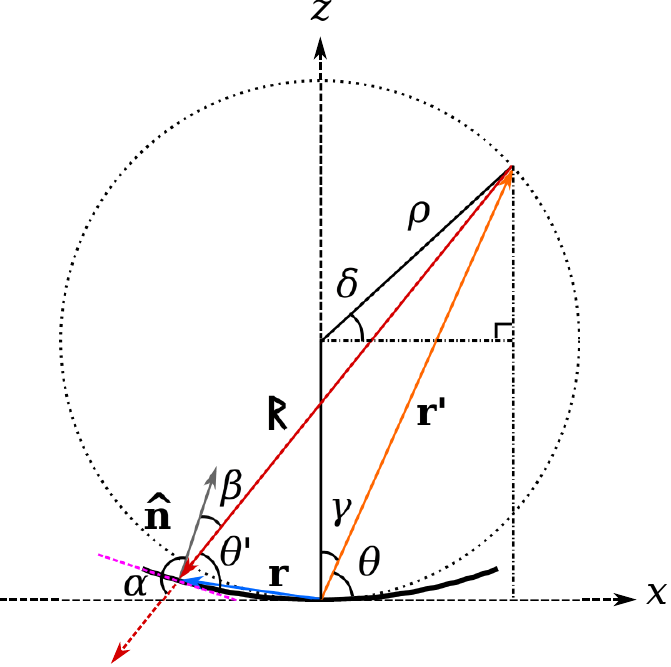}
\caption{Nomenclature used in the derivation of the Johann error.\label{fig:johann_error}}
\end{figure}
Consider a spherically bent crystal wafer with the meridional and sagittal bending radii $R_1$ and $R_2$, respectively. The surface of the spherical Johann-type analyser is approximately given by the constraint
\begin{equation}
f(x,y,z) = \frac{x^2}{2R_1}+\frac{y^2}{2R_2} -z  = 0,
\end{equation}
where $R$ is the bending radius. Let 
\begin{equation}
\mathbf{n} = - \nabla f = -\frac{x}{R_1} \hat{\mathbf{x}}
-\frac{y}{R_2} \hat{\mathbf{y}} + \hat{\mathbf{z}}.
\end{equation}
The surface normal vector field is thus $\hat{\mathbf{n}} = \mathbf{n}/n$, where
\begin{equation}
n = \sqrt{1 + \frac{x^2}{R_1^2} + \frac{y^2}{R_2^2}}
\end{equation}

Let us denote the distance from the source to the point $(x,y,z)$ on the crystal surface by the vector $\raidob$. According to Figure~\ref{fig:johann_error}, we find that $\raidob = \mathbf{r} - \mathbf{r}'$,
where $\mathbf{r}'$ is the position vector of the source and $\mathbf{r}$ is the position vector of the surface point in question. From Figure~\ref{fig:johann_error} we also see that
\begin{equation}
\mathbf{r}' = \rho \cos \delta \hat{\mathbf{x}}
+\rho(1 + \sin \delta) \hat{\mathbf{z}}.
\end{equation}
Since $\pi = \delta + \pi/2 + 2 \gamma$ and $\gamma = \pi/2 - \theta$, we find that $\delta = 2 \theta - \pi/2$. Thus
\begin{equation}
\mathbf{r}' = \rho \sin 2\theta \hat{\mathbf{x}}
+\rho(1 - \cos 2 \theta) \hat{\mathbf{z}}.
\end{equation}
Therefore
\begin{align}
\raidob &= (x - \rho \sin 2\theta ) \hat{\mathbf{x}} + y\hat{\mathbf{y}}
-\left(\rho(1 - \cos 2 \theta) - \frac{x^2}{2 R_1}-\frac{y^2}{2R_2} \right) \hat{\mathbf{z}} 
\label{eq:raido} \\
\Rightarrow |\raidob|^2 &= (x - \rho \sin 2\theta)^2 + y^2
+\left(\rho(1 - \cos 2 \theta)  - \frac{x^2}{2 R_1}-\frac{y^2}{2R_2} \right)^2 \nonumber \\
&=  \frac{1}{2} \left(x^2 +\frac{R_1}{R_2} y^2 - R_1^2   \right) \cos 2 \theta - xR_1 \sin 2 \theta 
 \nonumber \\ &+
\frac{1}{2} \left[x^2 + \left( 2 -  \frac{R_1}{R_2} \right)y^2 + R_1^2   \right]
+ \left(\frac{x^2}{2R_1} +\frac{y^2}{2R_2}\right)^2
\end{align}
where the fact that the  Rowland circle radius $\rho$ is half the meridional bending radius $R_1$. Since $\cos 2\theta = 1 - 2 \sin^2 \theta$ and $\sin 2\theta = 2\sin \theta \cos \theta$, we get
\begin{align}
|\raidob|^2 = R_1^2 \sin^2 \theta &\Bigg[ 1 + \frac{(R_2 - R_1) y^2}{R_2R_1^2 \sin^2 \theta}
- \frac{2x \cot \theta}{R_1} \nonumber \\ &+ \left(\frac{x^2}{R_1^2} + \frac{y^2}{R_1 R_2} \right) \cot^2 \theta   
 + \frac{1}{4 \sin^2 \theta} \left(\frac{x^2}{R_1^2} + \frac{y^2}{R_1 R_2}\right)^2 \Bigg] \nonumber \\
 \Rightarrow
\frac{1}{|\raidob|} = \frac{1}{R_1 \sin \theta} &\Bigg[ 1 + \frac{(R_2 - R_1) y^2}{R_2R_1^2 \sin^2 \theta}
- \frac{2x \cot \theta}{R_1} \nonumber \\ &+ \left(\frac{x^2}{R_1^2} + \frac{y^2}{R_1 R_2} \right) \cot^2 \theta   
 + \frac{1}{4 \sin^2 \theta} \left(\frac{x^2}{R_1^2} + \frac{y^2}{R_1 R_2}\right)^2 \Bigg]^{-1/2}
\end{align}

The cosine of angle $\alpha$ is now given by
\begin{equation}
\cos \alpha = \frac{\hat{\mathbf{n}} \cdot \raidob}{|\raidob|} = \frac{\mathbf{n} \cdot \raidob}{n |\raidob|}.
\end{equation}
Since
\begin{equation}
\mathbf{n} \cdot \raidob =
-\frac{x^2}{2 R_1} - \frac{y^2}{2R_2} +x \sin \theta \cos \theta -R_1 \sin^2 \theta
\end{equation}
we find that
\begin{align}
\cos \alpha &= - \sin \theta \left[1 - \frac{x}{R_1}\cot \theta + \frac{1}{2\sin^2 \theta}\left(\frac{x^2}{R_1^2} + \frac{y^2}{R_1 R_2} \right) \right]\left( 1 +\frac{x^2}{R_1^2} +\frac{y^2}{R_2^2}\right)^{-1/2}
\nonumber \\
&\times \left[1 - \frac{2x \cot \theta}{R_1} - \frac{(R_1 - R_2)y^2}{R_2 R_1^2 \sin^2 \theta} + \left(\frac{x^2}{R_1^2} + \frac{y^2}{R_1 R_2}\right) \cot^2 \theta + \frac{1}{4 \sin^2 \theta}\left(\frac{x^2}{R_1^2} + \frac{y^2}{R_1 R_2}\right)^2 \right]^{-1/2}.
\end{align}
Since $x/R$ and $y/R$ are small, we may expand $\cos \alpha$ as their series and retain only the terms up to the second order. Doing so we find
\begin{equation}
\cos \alpha \approx -\sin \theta - \frac{x^2}{2 R_1^2}\frac{\cos^2 \theta}{\sin \theta}
+ \frac{(R_1 -R_2)(R_1 \sin^2 \theta - R_2)}{2 R_1 R_2 \sin \theta}y^2.
\end{equation}
From Figure~\ref{fig:johann_error} we see that $\alpha + \beta = \pi$ and $\theta' + \beta = \pi/2$. Thus $\alpha = \pi/2 + \theta' \Rightarrow \cos \alpha = - \sin \theta'$  and
\begin{equation}
\sin \theta' = \sin \theta  +  \frac{x^2}{2 R_1^2} \frac{\cos^2 \theta}{\sin \theta}
- \frac{(R_1 -R_2)(R_1 \sin^2 \theta - R_2)}{2 R_1 R_2 \sin \theta}y^2. \label{eq:sin_theta_prime}
\end{equation}
By writing $\theta' = \theta + \Delta \theta$ and taking the first-order approximation $\sin\theta' \approx \sin \theta + \cos \theta \Delta \theta$, we find by comparing to Eq.~\eqref{eq:sin_theta_prime} that
\begin{equation}
\Delta \theta = \frac{x^2}{2 R_1^2} \cot \theta 
- \frac{(R_1 -R_2)(R_1 \sin^2 \theta - R_2)}{2 R_1 R_2 \sin \theta \cos \theta}y^2
\label{eq:johann_error_delta_theta}
\end{equation} 
Note that since Eq.~\eqref{eq:johann_error_delta_theta} is based on the first-order approximation of $\sin x$, it ceases to be valid near $\theta = \pi/2$ if $R_1 \neq R_2$. 

Alternatively, given in terms of energy the Johann error is
\begin{equation}
\Delta \mathcal{E} = \frac{hc}{2 d \sin \theta' } - \frac{hc}{2 d \sin \theta}
\approx -\frac{x^2}{2 R_1^2} \mathcal{E} \cot^2 \theta
+ \frac{(R_1 -R_2)(R_1 \sin^2 \theta - R_2)}{2 R_1 R_2 \sin^2 \theta} \mathcal{E}  y^2,
\label{eq:johann_error_delta_E}
\end{equation}
where $\mathcal{E} = hc/2 d \sin \theta$. Unlike Eq.~\eqref{eq:johann_error_delta_theta}, 
Eq.~\eqref{eq:johann_error_delta_E} is also valid at $\theta = \pi/2$ since we do not expand $\sin x$ with respect to its argument.

\referencelist[general_method_to_bent_crystal_wafers_references]

\end{document}